\documentclass{amsart}
\usepackage{amsthm} 
\usepackage{amstext}
\usepackage{amsbsy}
\usepackage{epsf}
\usepackage{mathrsfs}
\usepackage{amscd}
\usepackage{amsmath}
\usepackage{amsfonts}
\usepackage{hyperref}
\usepackage{latexsym}
\usepackage{amssymb}
\usepackage{fancyhdr}
\usepackage{times}
\usepackage{pst-tree}
\usepackage{textcomp}
\usepackage [T1] {fontenc}
\usepackage{graphicx}
\usepackage{pst-plot}
\usepackage[graphics,floats,textmath,displaymath,delayed]{preview}
\usepackage{url}

\newtheorem{theorem}{Theorem}
\newtheorem{lemma}{Lemma}
\newtheorem{proposition}{Proposition}

\theoremstyle{definition}

\theoremstyle{remark}
\newtheorem{remark}[theorem]{Remark}

\numberwithin{equation}{section}

\newcommand{\R}{\mathbb{R}}

\newcommand{\C}{\mathbb{C}}
\newcommand{\Z}{\mathbb{Z}}
\newcommand{\T}{\mathbb{T}}

\newcommand{\A}{\mathcal{A}}

\newcommand{\B}{\mathcal{B}}
\newcommand{\W}{\widetilde{W}}
\newcommand{\w}{\widetilde{w}}

\newcommand{\HH}{\mathcal{H}}

\newcommand{\LL}{\mathcal{L}}

\newcommand{\aalpha}{\alpha}
\newcommand{\ii}{\textrm{i}}

\newcommand{\WW}[1]{W^{#1}}
\newcommand{\VV}[1]{V^{#1}}

\newcommand{\x}{\bar{x}}
\newcommand{\X}{\bar{X}}

\begin{document}

\begin{center}

\normalsize{\textbf{A RENORMALIZATION PROOF OF THE KAM THEOREM FOR NON-ANALYTIC PERTURBATIONS}}

\vspace{1cm}

\footnotesize{

\textrm{EMILIANO DE SIMONE} \\

\vspace{0.2cm}

\textit{Department of Mathematics, University of Helsinki, P.O.Box 68 (Gustaf Hällströmin katu 2{b}) \\
Helsinki, 00014 , Finland \\ emiliano.desimone@helsinki.fi} 
}

\end{center}
\vspace{0.5cm}

\footnotesize{ 

\begin{center}
\parbox[c]{300pt}{We shall use a Renormalization Group (RG)
scheme in order to prove the classical KAM result in the case of a non-analytic
perturbation (the latter will be assumed to have continuous derivatives
up to a sufficiently large order).
We shall proceed by solving a sequence of problems in which the
perturbations are analytic approximations of the original one. We shall finally
show that the sequence of the approximate solutions will converge to a differentiable
solution of the original problem.}
\end{center}
\vspace{0.4cm}


\vspace{0.4cm}


}

\normalsize

\begin{section}{The KAM problem}

The first proof of the celebrated KAM theorem was presented by A.N. Kolmogorov 
in 1954 (see \cite{Kolmogorov1}). The theorem shows that if one adds a small perturbation to an integrable hamiltonian system, not all the invariant tori that foliate the phase space of the integrable system get destroyed. In fact, provided the perturbation is small, most of the phase space of the perturbed system is still occupied by invariant (though "distorted") tori. Kolmogorov's result was later improved by V.I.Arnold \cite{Arnold2, Arnold4} and J.Moser \cite{Moser1,Moser2}, the latter being the first to prove the KAM theorem in the case of a $\mathcal{C}^k$ perturbation.

The KAM theorem is strictly related to a well known perturbative series expansion, called the \textit{Lindsted Series}, whose convergence had troubled mathematicians since Poin\-ca\-ré's time. 
Even though Kolmogorov's, Arnold's and Moser's work indirectly showed that the Lindstedt series is convergent for analytic perturbation, it was only in 1988 that Eliasson, in \cite{Eliasson} proved it directly. By working on the series terms, Eliasson showed the mechanisms underlying the compensations that happen inside the series. Such compensations are shown to counter the effect of the huge contributions arising among the series terms due to the repeated occurrence of \textit{small denominators}.
Later on, Gallavotti, Chierchia, Gentile et al., noticed that Eliasson's method could be performed using the same diagrams that physicists had been using since Feynman. Namely one can represent the Fourier coefficients $\widehat{X}_k(q)$ of the terms of the Taylor expansion of the formal solution $\sum_k X_k\varepsilon^k$ (the Lindstetd series) by means of special Feynman diagrams: \textit{tree graphs}, i.e. ones without loops. The coefficient $\widehat{X}_k(q)$ will be given by a sum running over all tree graphs with $k$ vertices. The analogies between the methods used in Quantum Field Theory and Eliasson's proof of KAM were explained by the authors mentioned above in many influential papers (see for instance \cite{Chierchia1, Chierchia3, Gallalindt, Gallatwist, Gallatwist2, Gentile1}), where the convergence of the Lindstedt series for an analytic perturbation is proven by using a \textit{multiscale analysis}. One groups the "bad terms" \label{risonanze} (particular subgraphs called \textit{resonances}, which will be responsible for contributions inside $\widehat{X}_k(q)$ of the order $k!^s$ for $s>1$) that plague the Lindstedt series, into particular families inside which the diverging contributions compensate each other. Both the classical proof and the diagrammatic proof admit a  natural interpretation in terms of the \emph{Renormalization Group}(RG), see in particular \cite{Gallatwist2} which is explicitly based on a renormalization scheme inspired by the Wilsonian RG (For an explicit comparison between RG in QFT, in statistical mechanics and in classical mechanics see \cite{GallaRG}) . Furthermore, the compensations devised by Eliasson and later reinterpreted by Gallavotti in terms of diagrams can be shown to be identical to the so called \emph{Ward identities} of QFT, corresponding to a well known gauge symmetry.  

\textheight=7.7in

Making the latter interpretation explicit, J. Bricmont, K. Gaw\c edzki and A. Kupiainen in \cite{Antti1} gave yet another proof of the KAM theorem using the RG: here the small denominators are treated separately scale by scale, and the mechanism responsible for the compensations that make the Lindstedt series converge is explicitly shown to rely on the gauge symmetry expressed by the Ward identities.
Even though the Lindsted series converges only for analytic perturbations, the RG scheme used in \cite{Antti1} exploits a mechanism whose applicability is very general and not restricted to the convergence of such series. In the present paper, we use such mechanism in order to prove the classical KAM theorem in the case of a finitely many times differentiable perturbation, hence in a situation where the Lindsted series does not converge. In order to do that, and following Moser's original approach (see \cite{Moser1}), we solve a series of approximate problems obtained  by applying an ultraviolet cutoff to the perturbation. Furthermore the use of the Ward identities has to be slightly modified to fit the approximate scheme that we have to use due to the presence of a non-analytic perturbation. We shall see that our "modified identities", instead of implying that certain quantities vanish as in the original scheme \cite{Antti1}, will produce certain non zero terms which decay fast to zero and do not spoil the iteration. 
\end{section} 

\begin{section}{The Hamiltonian}
We restrict ourselves to study the Hamiltonian function of a system of rotators with a perturbation depending only on the angles (the treatment of the general case, where the perturbation depends on the actions as well, provides only heavier notation without shedding any further light on the proof):
\begin{align}
H(I,\theta) = \frac{I^2}{2} + \lambda V(\theta), \label{Hamiltonian}
\end{align}
where $\theta = (\theta_1, \ldots , \theta_d) \in \T^d$ are the angles describing the positions of the rotators and $I=(I_1, \ldots, I_d) \in \R^d$ are the conjugated actions. It generates the equations of motion
\begin{align}
\begin{cases} \label{motionpert}
\dot{I}(t) & = -\lambda \partial_{\theta}V(\theta(t)) \\
\dot{\theta}(t) & = I(t).
\end{cases}
\end{align}

When $\lambda = 0$ the trajectories are bound to run on the invariant tori $\T_{\omega} := \{ (\omega, \theta) \, | \, \theta \in \T^d \}$ and take the simple form
\begin{align}
\begin{cases} \label{motionnonpert}
I(t) & = \omega \equiv I(0) \\
\theta(t) & = \theta_0 + \omega t.
\end{cases}
\end{align}

When $\lambda >0$ the perturbation is "turned on", and we are interested in investigating the persistence of invariant tori and quasi-periodic solutions of \eqref{Hamiltonian}. We shall study such problem in the special case of a non-analytic perturbation $V$, the latter being assumed to be $\mathcal{C}^{\ell+1}$ for a sufficiently large integer $\ell$, whose size will be estimated later on. Namely the goal of this paper is the proof of the following classical result:
\begin{theorem}\label{KAM}
Let $H$ be the Hamiltonian \eqref{Hamiltonian}, with a perturbation $V$ such that its Fourier coefficients satisfy $\sum_q |q|^{\ell+1}|v(q)| \leq C$ (i.e. $\partial V \in \mathcal{C}^{\ell} $), and fix a frequency $\omega$ satisfying the diophantine property
\begin{align} 
|\omega \cdot q| \geq \gamma |q|^{-\nu} \quad \text{for some} \quad \gamma \in \R, \, \nu > d. \label{Diophantine}
\end{align}
Provided $|\lambda|$ is sufficiently small, if $\ell = \ell(\nu)$ is large enough, then for $s < \frac{2}{3}\ell -d$ there exists a  $\mathcal{C}^{s}$ embedding of the $d$-dimensional torus in $\T^d \times \R^d$, given by $\textrm{Id} + X_{\lambda}: \T^d \rightarrow \T^d \, , \, Y_{\lambda}: \T^d \rightarrow \R^d$, such that the solutions of the differential equation
\begin{align}
\dot\varphi = \omega 
\end{align}
are mapped into the solutions of the equations of motion generated by $H$, and the trajectories read
\begin{align}
\begin{cases}
\theta(t) & = \omega t + X_{\lambda}(\omega t) \\
I(t) & = Y_{\lambda}(\omega t),
\end{cases} \label{flow}
\end{align}
running quasi-periodically on a $d$-dimensional invariant torus with frequency $\omega$.
\end{theorem}

\begin{remark}
A crude estimate for $\ell(\nu)$ is given at page \pageref{oldstyle2} immediately after \eqref{oldstyle2}. Such a bad lower bound on $\ell$ shows that also our bound on $s$ is far from optimal: indeed in \cite{Chierchia2} and \cite{Salomon1} it is proved that, provided the perturbation is of class $\mathcal{C}^{\ell}$ for $\ell > \ell_0 = 4\nu +3$, then the embedding of the torus is of class $\mathcal{C}^{\ell-\ell_0}$. For $\ell$ as large as we require, it is always $s(\ell) < \ell - \ell_0$. 
\end{remark}

Plugging \eqref{flow} into the equations of motion \eqref{motionpert} we get a well known equation for $X$:
\begin{align}
\mathcal{D}^2 X(\theta) = - \lambda \partial_{\theta}V(\theta + X(\theta)), \quad \textrm{where} \quad \mathcal{D} := \omega \cdot \partial_{\theta}. \label{wellknown}
\end{align}
Trying to invert the operator $\mathcal{D}$ will lead us to deal with the infamous ``small denominators'': if we formally write the Fourier expression for $\mathcal{D}^{-1}$, the latter has the form $\frac{1}{(\omega \cdot q)}$, whose denominators can become arbitrarily small as $q$ varies in $\Z^d$.  As we shall see, a crucial role in controlling the size of such denominators will be played by the so-called \textit{Diophantine condition} \eqref{Diophantine}, which express the fact that $\omega$ cannot satisfy any resonance relation, not even approximately.
\end{section}

\begin{section}{The RG Scheme and the plan of the paper}\label{scheme}
As we already mentioned above, the main inspiration for the scheme used in this paper (and for most of the main techniques used) has been \cite{Antti1}, however we are in debt to \cite{Chierchia2} and \cite{Salomon1} for many fruitful ideas on how to adapt the proof to the case of a non-analytic perturbation, . 

From now on, we shall work with Fourier transforms, denoting by lower case letter the Fourier transform of functions of $\theta$, which will be denoted by capital letters:
\begin{align}
X(\theta) = \sum_{q \in \Z^d} e^{-\ii q\cdot\theta} x(q), \quad \textrm{where} \quad x(q) = \frac{1}{(2\pi)^d}\int_{\T^d} e^{\ii q\cdot \theta} X(\theta) d\theta.
\end{align}

In view of the  discussion at the end of the previous section, let us define
\begin{align}
W_0(X;\theta) := \lambda \partial_{\theta}V (\theta + X(\theta)).
\end{align} 
Denote by $G_0$ the operator $(-\mathcal{D}^2)^{-1}$ acting on $\R^d$-valued functions on $\T^d$ with zero average. In terms of Fourier transforms,
\begin{align}
(G_0 x)(q) = 
\begin{cases}
& \frac{x(q)}{(\omega \cdot q)^2} \quad \textrm{for} \quad  q \neq 0\\
& 0 \quad  \textrm{for} \quad q = 0;
\end{cases}
\end{align}
and we can write \eqref{wellknown} as the equations
\begin{align}
\begin{cases}
X = G_0 P W_0(X)  \label{Prob} \\
0 = \int_{\T^d} W_0(X;\theta) d\theta
\end{cases}
\end{align}
where $P$ projects out the constants: $PX = X - \int_{\T^d}X(\theta)d\theta$. \label{outtheconstants} 

Since we are not granted analyticity, we are not able to solve \eqref{Prob} by directly applying a renormalization scheme as in the case of an analytic perturbations (See for instance \cite{Antti1}); we shall instead proceed by means of analytic aproximations, which we know how to treat.
Let us set for $j= 1,2, \ldots$ the constants $\gamma_j, \, \alpha_j, \, \bar \alpha_j$ as follows 
\begin{align}
& \gamma_j := M8^j \notag \\
& \aalpha_j := \frac{1}{\gamma_{j-2}} = \frac{1}{M8^{j-2}} \notag \\
& \bar \alpha_j = \frac{1}{\gamma_{j+1}} \label{constants} 
\end{align}
where $M$ will be a large constant that we shall fix at the end of the proof. We define the analytic approximations
\begin{align}
\VV{j}(\xi) := \int_{\T^d} V(\theta)D_{\gamma_j}(\xi-\theta)d\theta = \sum_{|q|_{\infty}\leq \gamma_j} v(q)e^{iq\cdot\xi}. \label{Vaproximation}
\end{align}
where
\begin{align}
D_N(\theta) = \prod_{i=1}^d \frac{\sin{(N+\frac{1}{2})\theta_i}}{\sin{\frac{\theta_i}{2}}}
\end{align}
is the so-called Dirichlet Kernel 

With the latter setup, we get a sequence of ``analytically'' perturbed Hamiltonians:
\begin{align}
H(I,\theta) = \frac{I^2}{2} + \lambda \VV{j}(\theta), \label{AppHam}
\end{align}
givinge rise to a sequence of ``analytic'' problems 
\begin{align}
X(\theta) = G_0 P \WW{j}_0(X ; \theta) \label{AppProb}.
\end{align}
where 
\begin{align}
\WW{j}_0(X;\theta) \equiv \lambda \partial_{\theta}\VV{j} (\theta + X(\theta)) \label{First}
\end{align}
We shall first show how the renormalization scheme introduced in \cite{Antti1} can be employed to solve \eqref{AppProb} for a fixed $j$, and then modify the scheme to deal with all the $j$'s at once, and get a uniform upper bound for the coupling constant $0 < |\lambda|$.

We shall start our RG scheme in the same fashion as in \cite{Antti1}, and decompose 
\begin{align}
G_0 = G_1 + \Gamma_0
\end{align}
where $\Gamma_0$ will effectively involve only the Fourier components with $|\omega \cdot q|$ larger than $\mathcal{O}(1)$ and $G_1$ the ones with $|\omega \cdot q|$ smaller than that. Now we see that, if we write $X= Y + Z_0^j(Y)$, eq. \eqref{AppProb} becomes
\begin{align}
Y + Z_0^j(Y) = G_1PW_0^j(Y + Z_0^j(Y)) +\Gamma_0PW_0^j(Y + Z_0^j(Y)). \label{becomes0}
\end{align}
If $Z_0^j(Y)$ solves the large denominators problem, i.e.
\begin{align}
Z_0^j(Y) =  \Gamma_0P W_0^j(Y + Z_0^j(Y)) \label{zetasolves0}
\end{align}
one is left with the new effective problem
\begin{align} 
Y = G_1P W_1^j(Y) \quad \textrm{where} \quad W_1^j(Y) \equiv W_0^j(Y + Z_0^j(Y)). \label{newone0}
\end{align}
In order to exploit inductively the renormalization procedure above, we notice that Eq. \eqref{zetasolves0} is equivalent to the fixed point equation
\begin{align}
W^j_1(Y) = W^j_0(Y + \Gamma_0 W_1^j(Y)), \label{Larget10}
\end{align}
so that setting
\begin{align}
F^j_1(Y) \equiv Y + \Gamma_0W^j_1(Y)   \label{IndSol10}
\end{align}
the discussion above translates into the claim
\begin{align}
&  F^j_1(Y) \quad \textrm{is a solution to \eqref{AppProb}}  \iff \,  Y = G_1 P W^j_1(Y) \label{reduction10}
\end{align}
(see Eq. \eqref{becomes0}-\eqref{newone0}).
Thus \eqref{AppProb} reduces to the claim \eqref{reduction10} up to solving the easy large denominators problem \eqref{Larget10} and up to replacing the maps $W^j_0$ by $W^j_1$.

Suppose now that after $n-1$ inductive steps, the solution of Eq. \eqref{AppProb} is given by 
\begin{align}
F^j_{n-1}(Y) = Y + \Gamma_{n-2}W^j_{n-1}(Y)
\end{align}
where $Y$ must satisfy the equation
\begin{align}
Y = G_{n-1}P W^j_{n-1}(\X)\label{reduction0}
\end{align}
and $G_{n-1}$ contains only the denominators $|\omega \cdot q| \leq \mathcal{O}(\eta^n)$ where $0 < \eta \ll 1$ is fixed once for all.
The next inductive step consists of decomposing $G_{n-1} = G_n + \Gamma_{n-1}$ where $\Gamma_{n-1}$ involves $|\omega \cdot q|$ of order $\eta^n$ and $G_n$ the ones smaller than that.

If we define the maps $W^j_n(Y)$ as the solutions of the fixed point equation
\begin{align}
W^j_n(Y) = W^j_{n-1}(Y + \Gamma_{n-1}W^j_n(Y))  \label{IterationtW0},
\end{align}
and set
\begin{align}
F_n^j(Y) = F_{n-1}^j(Y + \Gamma_{n-1}W^{j}_n(Y)) \label{IterationF0},
\end{align}
we infer that $ F_n^j(Y)$ is the solution of \eqref{AppProb} if and only if $Y = G_nPW_n^j(Y)$, completing the following inductive step.
Finally it is easy to recover the inductive formulae
\begin{align}
& W^j_n(Y) = W_0^j(Y + \Gamma_{<n}W_n^j(Y))  \label{Titcumulative0} \\
& F_n^j(Y) = Y + \Gamma_{<n}W_n^j(Y), \label{Fcumulative0}
\end{align} 
where $\Gamma_{<n} \!\!=\!\! \sum_{k=0}^{n-1} \Gamma_k$.

Using \eqref{Titcumulative0} and \eqref{Fcumulative0} we see that, if $F_n^j(0)$ converges for $n \rightarrow \infty$ to $F^j$, we have
\begin{align}
F^j_n(0) & = \Gamma_{<n}W_n^j(0) \notag \\
& =  \Gamma_{<n}W_0^j(\Gamma_{<n}W_n^j(0)) \notag \\
& = \Gamma_{<n}W_0^j(F^j_n(0)) \label{mechanism} ,
\end{align}
and taking the limit for $n \rightarrow \infty$,
\begin{align}
F^j = G_0 W_0^j( F^j) \label{fsolution}
\end{align}
and $F^j$ is the solution of \eqref{AppProb} we were looking for.

This scheme, applied directly to the map $W^j_0$, would provide an approximate solution to \eqref{AppProb} for any fixed $j$, but that would not work, as either $|\lambda|$ or the set of allowed frequencies (labeling the preserved invariant tori), could shrink to zero as $j$ goes to infinity, making the procedure useless.
Instead we shall show that, by applying the above scheme to a slightly modified map $\W_0^j$, one can obtain a sequence of ``modified'' problems, the sum of whose solutions will converge to a $\mathcal{C}^{s}$ solution of our original problem, for all $s <\frac{2}{3}\ell - d$, provided $\ell$ is big enough and $|\lambda| \leq \lambda_0$.

We can assume inductively, as discussed earlier, that for $|\lambda| \leq \lambda_0$ and $k=0, \ldots j-1$ we have constructed real analytic functions $X_k(\theta)$ such that
\begin{align}
 X_{k}(\theta) = G_0P\WW{k}_0(X_{k};\theta) \label{AppSol},
\end{align}
From now on we shall write $\X := X_{j-1} = G_0\WW{j-1}_{0}(X_{j-1})$ and set 
\begin{align}
\W_0^j(Y) = W^j_0(\X + Y) - W^{j-1}_0 (\X).  \label{tildedef}
\end{align}
We notice that if the fixed point equation
\begin{align}
Y =  G_0\W_0^j(Y) \label{EquazDelta}
\end{align}
has a solution $Y_j$, then $X_j \equiv \X + Y_j$, is the solution to \eqref{AppProb} for $k=j$. Now we shall apply word by word the same scheme explained above, but to the map $\W_0^j(Y)$ instead of $W_0^j(Y)$. So one gets exactly the same iterative equations for the new effective problems and for the relative maps, with the only difference that we shall get a solution to a slightly different problem. Without repeating the mechanism of the scheme, which was explained above, we write the fundamental iterative equations for the new maps:
\begin{align}
\W^j_n(Y) = \W^j_{n-1}(Y + \Gamma_{n-1}\W^j_n(Y))  \label{IterationtW} \\
F_n^j(Y) = F_{n-1}^j(Y + \Gamma_{n-1}\W^{j}_n(Y)). \label{IterationF},
\end{align}
where with a slight abuse of notation, instead of writing $\widetilde F_n^j$, we used the same symbol $F_n^j$ as in \eqref{IterationF0} to denote a different map (see the discussion leading to \eqref{IterationtW0} and \eqref{IterationF0}). Let us remind that $ F_n^j(Y)$ is the solution of \eqref{EquazDelta}, if and only if $Y = G_nP\W_n^j(Y)$.
The inductive formulae \eqref{Titcumulative0} and \eqref{Fcumulative0} will obviously read
\begin{align}
& \W^j_n(Y) = \W_0^j(Y + \Gamma_{<n}\W_n^j(Y))  \label{Titcumulative} \\
& F_n^j(Y) = Y + \Gamma_{<n}\W_n^j(Y), \label{Fcumulative}
\end{align} 

Once again we shall use \eqref{Titcumulative} and \eqref{Fcumulative} to see that, if $F_n^j(0)$ converges for $n \rightarrow \infty$ to $F^j$, and get 
\begin{align}
F^j_n(0) = \Gamma_{<n}\W_0^j(F^j_n(0))  ,
\end{align}
which taking the limit for $n \rightarrow \infty$ will show that $F^j$ is the solution of \eqref{EquazDelta}.

At this point let us try to clarify the reason why we had to set an alternative RG scheme in order to deal with a non-analytic perturbation. For any fixed $j$ the "naive" iteration would give us a good control on the convergence (in $n$) of the renormalized solutions to the solution of the $j$-th problem, the latter turns out to be analytic in some $j$-dependent strip whose width shrinks to zero as $j$ grows. Thus increasing $j$ one eventually loses all the analyticity and using the bound obtained by the simple iteration one is not able to show the convergence to a smooth map for $j \rightarrow \infty$. On the other hand, using a "double" iterative procedure we do not look for bounds for the solution $X^j$ itself, instead we get certain bounds for the difference $Y^j := X^j - X^{j-1}$ between the solution of the $j$-th problem and the one of the $(j-1)$-th problems. In this manner the convergence of $X^j = X^{j-1} + Y^j$ can be recovered by looking at the telescopic series $\sum_j Y^j$. At the price of losing some of the regularity possessed by the original perturbation, the estimates recovered for $Y^j$ are able to counter the vanishing of analyticity, so that one still gets a smooth solution in the limit for $j \rightarrow \infty$. The details relative to this discussion are carried out in the final estimates of Section \ref{finalsection}.

The strategy of a "double" iteration in order to prove a $\mathcal C^k$ KAM theorem has been employed, in a somehow different fashion, in the papers \cite{Chierchia2} and \cite{Salomon1}. In these works the core of the analogy with the discussion above (and with our use of the modified scheme, that is), is best expressed by the Bernstein-Moser Theorem (see \cite{Chierchia2} p.47).

The rest of the paper is organized as follows. In section \ref{set} we define the functional spaces in which some preliminary bounds on the approximated perturbations $V^j$ are established. In Section \ref{cutt} we define the cutoff function $\Gamma$ and the functional spaces in which we want to solve our iterative scheme. In Section \ref{ndep} we prove some preliminary bounds on the maps $\w_n$ and $f_n$ which we are able to construct, using a Banach Fixed point Theorem, in a uniform complex neighborhood of $\T^d$ but only for $|\lambda| \leq \lambda_n$, where $\lambda_n$ is a sequence of positive real numbers converging to zero. 

In section \ref{WWW}, in order to extend the previously constructed maps to a uniform disk $|\lambda| \leq \lambda_0$ and obtain the required iteration, we introduce the Ward identities. To briefly explain how the latter enter the proof let us omit the indices $j$ and decompose $\w_n(y) = \w_n(0) + D\w_n(0)y +\delta_2w(y)$. We shall see that the linear term is the only problematic one for the iteration; in fact at each iterative step the constant part is easily dealt with by using the Diophantine condition plus the analyticity of $\w_n$, while the bounds on the high order terms follow easily since we are allowed to take $y$ as small as we want and to shrink the domain of analyticity by a small quantity at each step. 

The problem with the linear part arises when, while performing an iterative step, one has to invert the operator $\mathcal R := 1 - D\w_{n}\Gamma_{n}$. Such inversion is problematic since in the Fourier space $\Gamma_{n} \sim \mathcal{O}((\omega \cdot q)^{-2}) \sim \mathcal{O}(\eta^{-2n})$, which is obviously large. In the Fourier space the operator $D\w_n$, has a kernel which is a function of two momenta, label them $(q,q')$: on the off diagonal, i.e. where $q \neq q'$, such kernel can still be easily bounded using the Diophantine condition plus the analyticity, so as to compensate the large size of $\Gamma_{n}$ and allow for the inversion of $\mathcal R$. However, on the diagonal $q=q'$ the trick "Diophantine + analyticity" does not work anymore and we are bound to face the problem of \emph{resonances} \label{risonanze2}, i.e. terms in the Von Neumann series of $R$ consisting of repeated products of factors of the type $(\omega \cdot q)^{-2}$ which can become arbitrarily large ($\sim \mathcal{O}(\eta^{-2nk})$, for all $k$); this would irremediably spoil the iteration unless one is able to show that some compensations arise. 

At this point the Ward identities come into play and save the day. Expanding (in a sense that will be made more precise later) $D\w_n(q,q) = a_n + b_n(\omega \cdot q) + c_n(\omega \cdot q)^2 + \cdots$, our "revised" Ward identities allow one to conclude that the coefficients $a_n$ and $b_n$ decay to zero so fast as $n$ grows to $\infty$, that during the iteration one actually gets $Dw_n(q,q) \sim \mathcal{O}(\lambda (\omega \cdot q)^2)$, so that $D\w_n\Gamma_n \sim \lambda$ and $\mathcal R$ can be inverted. Note that in the analytic KAM proofs one gets from the "standard" Ward identities $a_n = b_n = 0$.

In Section \ref{mainprop} with the help of the identities we worked out in the previous Section, we prove Proposition \ref{main} which we refer to as the "main Proposition" since, as briefly discussed after its statement, it will naturally yield the main argument required in order to prove Theorem \ref{KAM}. 
In Section \ref{finalsection} we finally prove Theorem \ref{KAM} using Proposition \ref{main} and extending the functions $f_n$ constructed in Section \ref{ndep} to $|\lambda| \leq \lambda_0$.
In the Appendix the reader can find the proofs of the more technical results which, for the sake of brevity, have been omitted in the main sections.

\end{section}

\begin{section}{Setup and preliminary results}\label{set}


\begin{subsection}{Spaces}

From now on, for $q \in \Z^d$ we shall write $|q| := \sum_{i=1}^{d} |q_i|$ and we shall denote $\x \equiv x_{j-1}$ the inductive solution of the $(j-1)$-th analytic problem that we introduced in section \ref{scheme}, that is
\begin{align}
\x = G_0 w^{j-1}_0 (\x) \quad \x(0) = 0.
\end{align}

Recalling the definition \eqref{Vaproximation}, we write $\VV{j}(\theta) = \sum_q v^j(q) e^{iq\cdot \theta}$ by setting
\begin{align}
v^j(q) = \begin{cases} 
v(q) & \textrm{for} \: |q| \leq \gamma_j \\  
0 & \textrm{for} \: |q| > \gamma_j, \end{cases}
\end{align}

We shall denote 
\begin{align}
& \HH \equiv \{ (w(q))_{q\in \Z^d} \in \R^d  \: | \: \Vert w \Vert := \sum_q |w(q)| < \infty \} \\
& B(r) \equiv \{ w \in  \HH \: | \:  \Vert w \Vert \leq r  \}.
\end{align}
and let $H^{\infty}(B(r), \HH)$ denote the Banach space of analytic functions $w : B(r) \rightarrow \HH$ equipped with the supremum norm.

We shall assume inductively the following decay:
\begin{align}
& |\x(q)| \leq C \varepsilon A_j \frac{e^{-\frac{|q|}{4\gamma_j}}}{|q|^{\ell/3}} \quad \text{with} \quad A_j := \sum_{k=0}^{j-1}  \ell! \left( \frac{4}{M8^{k-5}} \right)^{\frac{\ell}{3}}  \label{xdecay},
\end{align}
where $M$ is as in \eqref{constants} and $\varepsilon \rightarrow 0$ \text{when} $|\lambda| \rightarrow 0$. The validity of the inductive bound \eqref{xdecay} at a first step $j_0$ can be easily recovered from the proof of the analytic KAM theorem; see, for instance, the bounds proven in \cite{Antti1},  Section $7$. 
\begin{remark}
Since we are not looking for optimal estimates, we shall leave the dependence of $\lambda$ on $\gamma$ implicit. However the classical KAM result in the non-degenerate case states that, fixed a  $\nu > d-1$, there exists a positive, $\gamma$-independent $\delta$ such that if $|\lambda| \leq \gamma^2 \delta$ then Theorem \ref{KAM} holds (see \cite{Pöschel2}).
  \end{remark}

From now on $C, C_1, C_2, C_3 \ldots$ will denote different constants which can vary from time to time. We can omit their dependence on the parameters when we think it is not important.
\end{subsection}


\begin{subsection}{A priori bounds for the approximate problems}
For $\sigma > 0$, denote by $\Xi_{\sigma}$ the complex strip
\begin{align}
\Xi_{\sigma} := \{\xi \in \C^d \, : \, |\textrm{Im}\xi| < \sigma\},
\end{align}
clearly there exists a $C >0$ such that for all $j$, the maps $V^j$ defined in \eqref{Vaproximation} obey
\begin{align}
\sup_{\xi \in \Xi_{\gamma_j^{-1}}} |V^j(\xi)| \leq C. \label{strips}
\end{align}
The bound \eqref{strips} implies the following 

\begin{lemma} \label{aprioridecay}
Let \eqref{xdecay} be valid for a suitably small $\varepsilon$ and write the Taylor expansion
\begin{align}
\partial V^j(\theta + \X(\theta) + Y) = \sum_{n=0}^{\infty} \frac{1}{n!}V^j_{n+1}(\theta + \X(\theta))(Y,\ldots,Y). 
\end{align} 
For each $|\sigma| < \frac{1}{5\gamma_j}$, there exists $ b > 0$, such that the coefficients $V^j_{n+1}(\theta + \X(\theta))$ belonging to the space of $n$-linear maps $\mathcal{L}(\C^{d}, \ldots , \C^{d}; \C^{d})$, have Fourier coefficients that decay according to the following bound
\begin{align}
\sum_{q \in \Z^d} e^{\sigma|q|} \Vert v_{n+1}^j(q;x)\Vert_{\mathcal{L}(\C^{d}, \ldots , \C^{d}; \C^{d})}  < b n! (2\gamma_j)^{n}.
\end{align}
\end{lemma}

The proof of the Lemma can be found in the Appendix  \ref{Appriori}


In view of the latter Lemma, let us introduce a translation $\tau_{\beta}$ by a vector $\beta \in \C^d$, $(\tau_{\beta}Y)(\theta) = Y(\theta - \beta)$. On $\HH$, $\tau_{\beta}$ is given by $(\tau_{\beta}y)(q) = y(q)e^{iq\cdot \beta}$. It induces a map $w \mapsto w_{\beta}$ from $H^{\infty}(B(r_0),\HH)$ to itself if we set
\begin{align}
w_{\beta}(y) = \tau_{\beta}(w(\tau_{-\beta}y)) \label{translation}
\end{align}  

The fixed-point equations, \eqref{IterationtW} and \eqref{Titcumulative} may be written in the form
\begin{align}
& \w^j_{n\beta}(y) = \w^j_{(n-1)\beta}(y + \Gamma_{n-1}\w_{n\beta}(y))\label{iterationt} \\
& \w^j_{n\beta}(y) = \w^j_{0\beta}(y + \Gamma_{<n}\w_{n\beta}(y))\label{titcumulative}
\end{align}

\begin{remark}
Note that, because of the definitions \eqref{tildedef} and \eqref{translation}, one has
\begin{align}
\w^j_{0\beta}(y) = \tau_{\beta}w_0^j (\x + \tau_{-\beta}y) - \tau_{\beta}w_0^{j-1}(\x) 
\end{align}
and the right hand side is \underline{not} $w_{0\beta}^j (\x + y) - w_{0\beta}^{j-1}(\x)$.
\end{remark}

Similarly, the equations \eqref{IterationF} and \eqref{Fcumulative} translate in the Fourier space to the relations
\begin{align}
& f^j_{n\beta}(y) = f^j_{(n-1)\beta}(y + \Gamma_{n-1}\w^j_{n\beta}(y)) \label{iterationf} \\
& f^j_{n\beta}(y) = y + \Gamma_{<n}\w_{n\beta}^j(y) \label{fcumulative}
\end{align}

\begin{proposition} \label{start}
Assume the inductive bound \eqref{xdecay} holds. Let $|\textrm{Im} \beta| < \frac{1}{8\gamma_j}$, and $\Vert y \Vert \leq \aalpha_j^{\frac{2}{3}\ell}$, see \eqref{constants}. We have 
\begin{align}
\sum_{q \in \Z^d}|\w^j_{0\beta}(y ; q)| \leq |\lambda|C_{d,\ell}\aalpha_j^{\frac{2}{3}\ell} \label{zero}
\end{align}
and furthermore, writing
\begin{align} 
\w^j_{0\beta}(y) = \w^j_{0\beta}(0) + D\w^j_{0\beta}(0)y + \delta_2 \w^j_{0\beta}(y), \label{sblit}
\end{align} 
where $D\w$ stands for the gradient of $\w$, we have
\begin{align}
& |\w^j_{0\beta}(0;q)| \leq C_1 |\lambda| \frac{\aalpha_j^{\frac{2}{3}\ell}}{|q|^{\frac{\ell}{3}}}  \label{uno} \\
& \Vert D\w^j_{0\beta}(0)y \Vert \leq C_2 |\lambda| \label{due}\\ 
& \Vert \delta_2 \w^j_{0\beta}(y) \Vert \leq C_3 |\lambda| \aalpha_j^{\ell}  \label{tre}
\end{align}

\end{proposition}

The proof of this Proposition can be found in the Appendix \ref{Appstart}

\end{subsection}


\begin{subsection}{Cauchy Estimates}
Let $h,h'$ be Banach spaces, we define $H^{\infty}(h ; h')$ as the space of analytic functions $w: h \rightarrow h'$ equipped with the supremum norm. Setting
\begin{equation}
\delta_k w(x) = w(x) - \sum_{s=0}^{k-1} \frac{1}{s!} D^s w(0)(y),
\end{equation}
we shall make use of the following Cauchy estimates throughout the paper: 
\begin{align}
& \sup_{\Vert y \Vert \leq r - \delta} \Vert Dw(y) \Vert \leq \sup_{\Vert y \Vert \leq r} \frac{1}{\delta} \Vert w(y) \Vert \label{Cauchy1}\\
& \sup_{\Vert y \Vert \leq r'\mu} \Vert \delta_k w(y) \Vert \leq \frac{\mu^k}{1-\mu}\sup_{\Vert y \Vert \leq r'} \Vert w(y) \label{Cauchy2} \Vert
\end{align}
Furthermore we shall also need the following estimate: let $w_i \in H^{\infty}(B(r) \subset h \; ; h')$ for $i=1,2$, and $w \in H^{\infty}(B(r') \subset h' \; ; h'')$, then, if $\sup_{\Vert y \Vert_h \leq r} \Vert w_i (y) \Vert_{h'} \leq \frac{1}{2}r'$, we have
\begin{align}
\sup_{\Vert y \Vert_h \leq r} \!\!\!\! \Vert w\circ w_1(y) - w\circ w_2(y)  \Vert_{h''} \leq \frac{2}{r'} \!\!\sup_{\Vert y' \Vert_{h'} \leq r'} \!\!\!\!\! \Vert w(y') \Vert_{h''} \!\! \sup_{\Vert y \Vert_h \leq r} \!\!\!\! \Vert w_1(y) - w_2(y) \Vert_{h'} \label{Cauchy3}
\end{align}  
\end{subsection}

\end{section}
\begin{section}{The Cutoff and n-dependent spaces}\label{cutt}
To define the operators $\Gamma_n$ - that establish our renormalization - we will divide the real axis in scales. We shall fix $\eta \ll 1$ (once and for all) and introduce the so-called "standard mollifier" by 
\begin{equation} h(\kappa) =
\begin{cases}
C_{\kappa}e^{\frac{1}{\kappa^2-1}} \quad &\textrm{if} \; |\kappa| < 1 \\
0 \quad &\textrm{if} \; |\kappa| \geq 1
\end{cases}  
\end{equation}
with the constant $C$ chosen such that $\int_{\R}h dx =1$. Now let us define $ \bar\chi \in \mathcal{C}^{\infty}(\R)$ by
\begin{align}
\bar\chi(\kappa) := 1-\frac{2}{1-\eta}\int_{\frac{1+\eta}{2}}^{\infty} h\left(\frac{2(|\kappa|-y)}{1-\eta}\right)dy  
\end{align}
so that 
\begin{equation} \bar\chi(\kappa) =
\begin{cases}
1 \quad &\textrm{if} \; |\kappa| < \eta \\
0 \quad &\textrm{if} \; |\kappa| \geq 1
\end{cases}  
\end{equation}
and
\begin{align}
\sup_{\kappa \in \R}|\partial_{\kappa}\bar\chi(\kappa)| \: , \: \sup_{\kappa \in \R}|\partial^2_{\kappa}\bar\chi(\kappa)| \leq C,  \label{uniform}
\end{align}
for a suitable constant $C$.
We set
\begin{align}
\bar{\chi}_n(\kappa) = \bar{\chi}(\eta^{-n}\kappa)
\end{align}
and set
\begin{align}
& \chi_0(\kappa) = 1 - \bar{\chi}_1(\kappa) \notag \\
& \chi_n(\kappa) = \bar{\chi}_{n}(\kappa) - \bar{\chi}_{n+1}(\kappa) \quad \textrm{for} \quad n \geq 1.
\end{align}

Finally we define the diagonal operator $\Gamma_n : \HH \rightarrow \HH$
\begin{align}
\Gamma_n(q , q') = \frac{\chi_n(\omega \cdot q)}{(\omega \cdot q)^2}\delta_{q,q'} := \gamma_n(\omega \cdot q)\delta_{q,q'},
\end{align}
so that $\left\{ q \, : \, \Gamma_{n-1}(q,q) \neq 0 \right\} = \{ \eta^{n+1} \leq |\omega \cdot q| \leq \eta^{n-1} \}$.
The formulae coming from our renormalization scheme suggest us to define $n$-dependent norms and spaces. For $n \geq 2$ we set the quotient space $\HH_{-n} := h / \overset{-n}{\sim}$, where $h$ is the space of functions $\{ w : \Z^d \rightarrow \C^d \}$ and $\overset{-n}{\sim}$ is the equivalence relation in $h$ defined by
\begin{align}
w \overset{-n}{\sim} w' \Longleftrightarrow w(q) = w'(q) \quad \text{for all} \quad |\omega \cdot q| \leq \eta^{n-1}
\end{align}
We make $\HH_{-n}$ Banach spaces by endowing them with the norms
\begin{align}
\Vert w \Vert_{-n} = \sum_{|\omega \cdot q| \leq \eta^{n-1}} |w(q)|.
\end{align}
\begin{remark}
It might be useful to stress that when we shall write an equality $w = w'$ between two elements of $\HH_{-n}$, that means by definition $w \overset{-n}{\sim} w'$; hence as functions over $\Z^d$, $w$ and $w'$ coincide only on the set   $\{ q : |\omega \cdot q| \leq \eta^{n-1}\}$     
\end{remark}

Next we consider the projection
\begin{align}
P_n(y)(q) =
\begin{cases}
y(q) \quad & \textrm{if} \, |\omega \cdot q| \leq  \eta^{n-1} \\
0 \quad & \textrm{otherwise}.
\end{cases}
\end{align}
and define the spaces
\begin{align}
\HH_n \equiv P_n \HH,  
\end{align}
equipped with the norm inherited from $\HH$:
\begin{align} 
\Vert y \Vert \equiv \sum_q |y(q)| = \sum_{|\omega \cdot q| \leq \eta^{n-1}} |y(q)|,
\end{align}
\begin{remark} 
For $y \in \HH_n $,  $\Vert y \Vert = \Vert y \Vert_{-n}$, even though in general $\Vert \cdot \Vert \neq \Vert \cdot \Vert_{-n}$,
\end{remark}
Note the natural embeddings for $n \geq 2$:
\begin{align}
\HH_n \rightarrow \HH_{n-1} \rightarrow  \HH \rightarrow \HH_{-n+1} \rightarrow \HH_{-n} \label{embeddings}
\end{align}
We shall denote by $B^j_n(r)$ the open ball in $\HH_n$ of radius $r_j$.

If we define the cutoff with ``shifted kernel''
\begin{align}
\Gamma_n[\kappa](q) =  \gamma_n(\omega \cdot q + \kappa) \label{Gammashifted}
\end{align}   
we can prove the following:
\begin{lemma}\label{cutoff}
For $i=0,1,2$ and $|\kappa| \leq \eta^n$, the cutoff functions obey the following estimates
\begin{align}
\Vert \partial^i_{\kappa} \Gamma_{n-1}[\kappa]\Vert \leq C\eta^{-(2+i)n}
\end{align}
\end{lemma}
\begin{proof}
The proof is trivial, since for $\tilde \kappa = \kappa + \omega \cdot q$ we have, by definition, 
\begin{align}
\Gamma_{n-1}[\kappa](q) = \chi_{n-1}(\tilde \kappa)/\tilde\kappa^2  
\end{align} 
and $\chi_{n-1}(\kappa) = 0$ for $|\kappa| \leq \eta^{n-1}$.
\end{proof}
\end{section}


\begin{section}{n-dependent bounds}\label{ndep}
Our final goal is to show that the maps $\w_n^j$ and $f_n^j$ exist for all $j$ and $n$, provided $\lambda$ is small enough in an $n$-independent way. However for later purposes it will be useful to show first some simple $n$-dependent bounds, uniform in $\beta$. Such bounds are carried out quite easily in the next proposition:
\begin{proposition} \label{easy}
For any sufficiently small $r > 0$, there exists a sequence of positive numbers  $\lambda_n \overset{n \rightarrow \infty}{\longrightarrow} 0$ such that for $|\lambda| \leq \lambda_n$ and $|\textrm{Im} \, \beta| \leq \aalpha_j/2$ the equations \eqref{titcumulative} have a unique solution $\w_n^j \in H^{\infty}(B(\aalpha_j^{\frac{2}{3}\ell}r^n) , \HH)$ with
\begin{align}
\sup_{y \in B(\aalpha_j^{\frac{2}{3}\ell}r^n)} \Vert \w_n^j \Vert \leq C_{d,\ell} \alpha_j^{\frac{2}{3}\ell} |\lambda| \label{go}
\end{align}
where $C_{d,\ell}$ is as in Proposition \ref{start}. Furthermore the maps $f_{n\beta}^j$ defined by Eqs. \eqref{fcumulative} belong to $H^{\infty}(B(\aalpha_j^{\frac{2}{3}\ell} r^n),\HH)$. They satisfy the bounds 
\begin{align}
\sup_{\Vert y \Vert \leq \aalpha_j^{\frac{2}{3}\ell}r^n} \Vert f_{n\beta}^j(y) \Vert \leq 2\aalpha_j^{\frac{2}{3}\ell} r^n.
\end{align}
Moreover, $w_{n\beta}^j$ and $f_{n\beta}^j$ are analytic in $\lambda$ and $\beta$ and they satisfy the recursive relations \eqref{iterationt} and \eqref{iterationf}, respectively.
\end{proposition}

Deferring the proof of Proposition \ref{easy} to the Appendix \ref{Appeasy} let us point out that the maps $\w_{n\beta}$ that have been constructed in Proposition \ref{easy} only for $|\lambda| \leq \lambda_n$ will be analytically extended to a uniform disk $|\lambda| \leq \lambda_0$  in Proposition \ref{main}, provided that, as $n$ increases, we shrink the strip in which $\beta$ is allowed to vary. However we shall make use of the bounds obtained in Proposition \ref{easy} for the maps $f_n^j$ (which shall be analytically extended to  $|\lambda| \leq \lambda_0$ as well) in the beginning of Section \ref{finalsection}.

\end{section}

\begin{section}{The Ward identities (revised)} \label{WWW}
\begin{subsection}{The identities}
We shall prove in this section some properties of the maps $w^j_n$ which will be essential in the proof of the main Proposition. As mentioned in Section \ref{scheme}, they will come into play when trying to bound the diagonal part of the kernel of $D\w_n(y)$. Such bounds will show that compensations happen among the so-called \textit{resonances}, the latter being the particular terms that make the convergence of the Lindstedt series (see p.\pageref{risonanze} and p.\pageref{risonanze2}) problematic. For the definition of ``resonances'' see also \cite{Chierchia3, Eliasson, Gallalindt, Gallatwist, Gallatwist2}.
We shall work out some identities which can be considered as "revised" Ward identities (for the "standard" Ward identities see \cite{Antti1, Eliasson, Gallatwist2, GallaRG}) for the maps $\w^j_n$ that we constructed in Proposition \ref{easy}. We shall omit the indeces $j$, writing $X= \X$, $V = V_j$, $\widehat{V} = V_{j-1}$, $W = W^j$ and $U= W^{j-1}$, and the summations over repeated indeces will be understood. The basic identity reads
\begin{align}
& \int_{\T^d}\W_n^{\gamma}(Y; \theta)d\theta = \int_{\T^d} Y^{\alpha}(\theta)\partial_{\gamma}W_0^{\alpha}(X+Y+\Gamma_{<n}\W_n(Y)  ; \theta)d\theta \notag \\
& \quad + \int_{\T^d} G_n U_0^{\alpha}(X ; \theta)\partial_{\gamma}\W_n^{\alpha}(Y ; \theta)d\theta, \label{basic} 
\end{align}
whose proof is given in the Appendix \ref{AppWard}.
Once \eqref{basic} is proven, we can transpose it into the Fourier space language:
\begin{align}
\w_n^{\gamma}(y;0) & = -\sum_{q \neq 0}iq^{\gamma}y^{\alpha}(q)w_0^{\alpha}(x+y+\Gamma_{<n}\w_n(y);-q) \notag \\
& \quad -\sum_{q \neq 0}iq^{\gamma} \bar{\chi}_n(\omega \cdot q)\x^{\alpha}(q)\w_n^{\alpha}(y;-q) \label{fbasic},
\end{align}
so it immediately follows that
\begin{align}
\w_n^{\gamma}(0;0) = -\sum_{q \neq 0}iq^{\gamma}\bar{\chi}_n(\omega \cdot q)\x^{\alpha}(q)\w_n^{\alpha}(0;-q). \label{Ward1}
\end{align}
Differentiating \eqref{fbasic} with respect to $y^{\alpha}(q)$ and evaluating it at $y=0$, we get 
\begin{align}
\frac{\partial \w_n^{\gamma}(y;0)}{\partial y^{\alpha}(q)}\Big|_{y=0} & = - iq^{\gamma}w_0^{\alpha}(x+\Gamma_{<n}\w_n(0);-q) \notag \\
& \quad - \sum_{q'\neq 0}iq'^{\gamma} \bar{\chi}_n(\omega \cdot q')\x^{\beta}(q')\frac{\partial \w_n^{\beta}(y;q')}{\partial y^{\alpha}(q) }\Big|_{y=0} \label{Ward2}
\end{align}

\end{subsection}

\begin{subsection}{An interpolation of the linear term's kernel} \label{resonances}
To use the identities we obtained in the last section we have to introduce smooth interpolations of the kernels of the maps $D\w_n$, constructed in Proposition \ref{easy} for $|\lambda| \leq \lambda_n$. Such interpolations will prove essential in order to exhibit the compensations that occur inside the diagonal part of the kernel $D\w_n$ (among the "resonances") . Differentiating \eqref{titcumulative} we get
\begin{align}
D\w_{n\beta}(y) = \left[ 1 - D\w_{0\beta}(y_n)\Gamma_{<n} \right]^{-1}D\w_{0\beta}(y_n) \quad \textrm{with} \quad y_n \equiv y + \Gamma_{<n}\w_{n\beta}(y). \label{dereasy}
\end{align} 
We will show that the diagonal part of the kernel $D\w_{n\beta}(y;q,q)$ depends on $q$ only through $\omega \cdot q$. In order to show this, for $p \in \Z^d$, let $t_p : \LL(\HH; \HH) \rightarrow \LL(\HH; \HH)$ be the continuous automorphism that maps $a \in \LL(\HH; \HH)$ into $t_p a \in \LL(\HH; \HH)$:
\begin{align}
(t_p a) (q,q') = a(q+p, q'+p),
\end{align}
that is, $t_p$ shifts the kernel of the operator $a$ by $p$. For $n=0$ we have that $t_p D\w^j_{0\beta} = D\w^j_{0\beta}$ for all $p \in \Z^d$, since the kernel $D\w^j_{0\beta}(y;q,q')$ is function of $q-q'$ only. The latter observation and the definition \eqref{Gammashifted} allow us to conclude that, applying $t_p$ to \eqref{dereasy}, we get
\begin{align}
t_p D\w^j_{n\beta}(y) = \left[ 1 - D\w^j_{0\beta}(y_n)\Gamma_{<n}(\omega \cdot p) \right]^{-1}D\w^j_{0\beta}(y_n),
\end{align}
showing that $t_p D\w^j_{n\beta}(y)$ depends on $p$ only through $\omega \cdot p$. Therefore we can define a smooth interpolation of  $t_p D\w^j_{n\beta}(y)$ in the following way: denote $\pi^j_{0\beta}(y) = D\w_{0\beta}(y)$ and define for $n \geq 1$ and $|\kappa| \leq \eta^n$,
\begin{align}
\pi^j_{n\beta}(\kappa;y) = \left[ 1 - \pi^j_{0\beta}(y_n)\Gamma_{<n}(\kappa) \right]^{-1}\pi^j_{0\beta}(y_n). \label{picumulative}
\end{align}
Inequality \eqref{virtue} shows that for $y \in B(\aalpha_j^{\frac{2}{3}\ell}r^n) \subset \HH$, $\Vert y_n \Vert \leq \frac{1}{2}{\aalpha_j^{\frac{2}{3}\ell}}$, so Proposition \ref{start} and the Cauchy estimate \eqref{Cauchy1} imply for such $y$
\begin{align}
& \Vert \pi^j_{0\beta}(y_n) \Vert_{\LL(\HH;\HH)} \leq \sup_{\Vert y \Vert \leq \frac{1}{2}\aalpha_j^{\frac{2}{3}\ell}} \Vert D\w^j_{0\beta}(y) \Vert_{\LL(\HH;\HH)} \notag \\
& \leq \frac{2}{\aalpha_j^{\frac{2}{3}\ell}} \sup_{\Vert y \Vert \leq {\aalpha_j^{\frac{2}{3}\ell}}} \Vert \w^j_{0\beta}(y) \Vert \leq |\lambda| 2 C_{d,\ell}.
\end{align}
The latter discussion implies that $\pi^j_{n\beta}(\kappa;y)$ is analytic for $|\lambda| \leq \lambda_n$, $|\textrm{Im}\,\beta| < \bar \alpha_j$, $y \in B(\aalpha_j^{\frac{2}{3}\ell}r^n) \subset \HH$,  and $\mathcal{C}^{\infty}$ for $|\kappa| \leq \eta^n$  with norm, say,
\begin{align}
\Vert \pi_{n\beta}^j(\kappa;y) \Vert_{\LL(\HH;\HH)} \leq \sqrt{|\lambda|}.
\end{align}
Furthermore $\pi^j_{n\beta}(\kappa;y)$ is a smooth interpolation of the kernel of $t_pD\w_{n\beta}(y)$, meaning that for $p \in \Z^d$
\begin{align}
t_pD\w_{n\beta}(y) = \pi^j_{n\beta}(\omega \cdot q ;y). 
\end{align} 
Differentiating Eq. \eqref{picumulative} with respect to $\kappa$ we get the useful identity
\begin{align}\partial_{\kappa} \pi^j_{n\beta}(\kappa ;y) =\pi^j_{n\beta}(\kappa ;y) \partial_{\kappa}\Gamma_{<n}(\kappa)\pi_{n\beta}(\kappa;y). \label{piderivative}
\end{align} 
For $\Vert y \Vert \leq \alpha_j^{\frac{2}{3}\ell}r^n$ and $|\kappa| \leq \eta^n$ the following recursive relation holds:
\begin{align}
\pi^j_{n\beta}(\kappa ;y) = \left[1 -  \pi^j_{(n-1)\beta}(\kappa ;\tilde y) \Gamma_{n-1}(\kappa) \right]^{-1} \pi^j_{(n-1)\beta}(\kappa ;\tilde y) \label{pirecursive}
\end{align}
where $\tilde y = y + \Gamma_{n-1}\w_{n\beta}(y)$.

\end{subsection}
\end{section}

\begin{section}{The Main Proposition}\label{mainprop}
To simplify the notations, we shall denote by $B_n^j$ the open ball in $\HH_n^j$ of radius $\aalpha_j^{\frac{2}{3}\ell}r^n$ and by $\A_n^j$ the space $H^{\infty}( B_n^j, \HH_{-n})$. Finally $\Gamma$ will stand for $\Gamma_{n-1}$.
\begin{proposition} \label{main}
\begin{itemize}
\item[\textbf{(a)}]
There exist positive constants $r_j$,  $\lambda_0$, 
and $\bar \alpha_{j,n}$ where
\begin{align}
\bar \alpha_{(j;n)} = \frac{n+2}{2n+2}\bar \alpha_j \quad n \geq 1, \label{sequence}
\end{align}
such that, for $|\text{Im} \beta| \leq \alpha_{(j;n)}$ and $|\lambda| \leq |\lambda_0|$ there exist solutions $\w_{n\beta}^j \equiv \w_n^j$ of Eqs. \eqref{iterationt} such that $\w_n^j$ belong to $\A_n^j$, and are analytic in $\lambda$.
\item[\textbf{(b)}]
Writing
\begin{align}
\w_n^j(y) = \w_n^j(0) + D\w_n^j(0)y + \delta_2 \w_n^j(y) \label{Derrida}
\end{align}
we have
\begin{align}
& |\w_n^j(0;q)|   \leq \varepsilon \left( 2^{n+1}-1 \right)\frac{\aalpha_j^{\frac{2}{3}\ell}}{|q|^{\frac{\ell}{3}}} \quad \text{for} \quad 0 < |\omega \cdot q| \leq \eta^{n-1} \label{primo}\\
& \Vert \delta_2\w_n^j \Vert_{\A_n^j}  \leq \varepsilon\aalpha_j^{\ell} r^{\frac{3}{2}n} \label{terzo}
\end{align}
where $\varepsilon \rightarrow 0$ as $\lambda \rightarrow 0$.
\item[\textbf{(c)}] Furthermore \label{etichetta}
\begin{align}
\Vert D\w_n^j(y) \Vert_{\LL(n;-n)} \leq \varepsilon \eta^{2n} \label{secondo}
\end{align}
\end{itemize}
\end{proposition}

\begin{remark}
The reason why Proposition \ref{main} is our "main" Proposition lies in the mechanism of our RG scheme and, at this point, it can be understood by looking at the formal manipulation in Section \ref{scheme} starting with Eq. \eqref{mechanism} (mutatis mutandis, i.e. changing $W_n^j \rightarrow \W_n^j$). Provided $\w_n$ exists and it is as small as yielded by Proposition \ref{main} (see in particular the bound \eqref{oldstyle} directly implied by \eqref{primo}), then, using the recursion \eqref{iterationf}, the sequence $f_n^j(0) = \Gamma_{<n} \w_n^j(0)$ can be proved to be Cauchy and to converge to the required solution of \eqref{EquazDelta} (see Eq. \eqref{fsolution}). For the technical details on how Proposition \ref{main} implies Theorem \ref{KAM} see Section \ref{finalsection}.
\end{remark}


\begin{subsection}{Proof of (a)}
First of all, we show that \eqref{primo} implies for all $n \geq 1$:
\begin{align}
\Vert P\w^j_n(0) \Vert_{-n} \equiv \sum_{|\omega \cdot q| \leq \eta^{n-1}} |\w^j_n(0 ; q)| \leq \varepsilon \aalpha_j^{\frac{2}{3}\ell} r^{2n} \label{oldstyle} 
\end{align}
In fact the diophantine condition \label{discussion} \eqref{Diophantine}, forces the sum defining the norm to be taken over $q$ such that $|q| \geq \gamma^{\frac{1}{\nu}} \eta^{-\frac{n-1}{\nu}} \,$, 
hence we can estimate
\begin{align}
& \sum_{|\omega \cdot q| \leq \eta^{n-1}}|w_{n}(0;q)| \leq  \sum_{|q| \geq  \gamma^{\frac{1}{\nu}} \eta^{-\frac{n-1}{\nu}} }|w_{n}(0;q)| \notag \\
& \leq \varepsilon \left ( 2^{n+1}-1\right ) \aalpha_j^{\frac{2}{3}\ell}\sum_{|q| \geq  \gamma^{\frac{1}{\nu}} \eta^{-\frac{n-1}{\nu}} }  \frac{1}{|q|^{\frac{\ell}{3}}} \notag \\
& \leq \varepsilon \gamma^{\frac{d-\frac{\ell}{3}}{\nu}}\aalpha_j^{\frac{2}{3}\ell}  \left ( 2^{n+1}-1\right )  \eta^{\frac{n-1}{\nu}(\frac{\ell}{3}-d)} \notag \\
& \leq \varepsilon \aalpha_j^{\frac{2}{3}\ell} r^{2n} \label{oldstyle2}
\end{align}
for $\varepsilon = \varepsilon(d, \gamma, \nu)$ and $\ell \geq 12\nu \log_{\eta}(r/2) +3d$.


\begin{remark}
Note that \eqref{oldstyle} can be trivially improved with
\begin{align}
\Vert \w^j_m(0) \Vert_{-n} \leq \varepsilon \aalpha_j^{\frac{2}{3}\ell} r^{2n}
\end{align}
for all $m \leq n$. Anyway we shall not need the latter bound and in the following we shall always use \eqref{oldstyle}. 
\end{remark}

Consider now the equation \eqref{iterationt}. The decomposition \eqref{Derrida} implies
\begin{align}
\w^j_n(y) & = \w^j_{n-1}(0) + D\w^j_{n-1}(0)(y + \Gamma \w^j_n(y)) + \delta_2 \w^j_{n-1}(y + \Gamma \w^j_n(y)) 
\end{align} 
from which, setting $H \!= \!(1 - D\w^j_{n-1}(0)\Gamma)^{-1}$ and $\widetilde{H} \!=\! 1 + \Gamma HD\w^j_{n-1}(0) \!=\! (1-\Gamma D\w^j_{n-1}(0))^{-1}$, we deduce that
\begin{align}
\w_n^j(y) & =  H\w_{n-1}^j(0) +  HD\w^j_{n-1}(0)y + u(y) \label{decomposition}
\end{align}
where
\begin{align}
u(y) & = H \delta_2 \w^j_{n-1}(y + \Gamma \w^j_n(y)) = H \delta_2 \w^j_{n-1} (\Gamma H\w^j_{n-1}(0) + \widetilde{H}y + \Gamma u(y)) \label{u}
\end{align}
The bound \eqref{secondo} with $n$ replaced by $n-1$, together with Lemma \ref{cutoff} and the definition of the norms imply
\begin{align}
\Vert H \Vert_{\LL(-n+1;-n+1)} \: , \:  \Vert \widetilde{H} \Vert_{\LL(-n+1;n-1)} \leq 1 + C\varepsilon \leq 2, \label{H} 
\end{align}
for $|\lambda|$ small enough.

To solve Eq. \eqref{u} we use the Banach Fixed Point Theorem. Once $u$ is given, we can recover the existence of $\w^j_n$ solving \eqref{decomposition}. The solution of \eqref{u} can be given as the fixed point of the map $\mathcal{G}$ defined by
\begin{align}
\mathcal{G}(u) =  H \delta_2 \w^j_{n-1} (\tilde{y}) \quad \textrm{with} \quad \tilde{y} = \Gamma H\w^j_{n-1}(0) + \widetilde{H}y + \Gamma u(y).
\end{align} 
We shall show that $\mathcal{G}$ is a contraction in the ball
\begin{align}
\mathcal{B}^j = \{ u \in H^{\infty}(B_{n-1}^{j,\delta}, \HH_{-n+1}) \, | \, \Vert u \Vert_{\mathcal{B}^j} \equiv \sup_{y \in  B_{n-1}^{j,\delta}} \Vert u(y) \Vert_{-n+1} \leq 2\varepsilon \aalpha_j^{\ell}r^{\frac{3}{2}(n-1)} \} \label{ball},
\end{align}
where $B_{n-1}^{j,\delta} \subset \HH_{n-1}$ is the open ball of radius $\aalpha_j^{\frac{2}{3}\ell}r^{n-\delta}$ for $0 \leq \delta < 1$ and $r_j = r_j(\delta)$.
Indeed, for $y \in \HH_{n-1}$ such that $\Vert y \Vert_{n-1} \leq \aalpha_j^{\frac{2}{3}\ell} r^{n-\delta}$, we get $\tilde{y} \in \HH_{n-1}$ with
\begin{align}
\Vert \tilde{y} \Vert_{n-1} & \leq 2C\eta^{-2n}\Vert w_{(n-1)\beta}(0) \Vert_{-n+1} + 2\aalpha_j^{\frac{2}{3}\ell}r^{n-\delta} + 2C\eta^{-2n}\varepsilon \aalpha_j^{\ell}r^{\frac{3}{2}(n-1)}  \notag \\
& \leq 2C\eta^{-2n}\varepsilon \aalpha_j^{\frac{2}{3}\ell} r^{2(n-1)} + 2\aalpha_j^{\frac{2}{3}\ell}r^{n-\delta} + 2C\eta^{-2n}\varepsilon \aalpha_j^{\ell}r^{\frac{3}{2}(n-1)} \notag \\
& \leq \frac{1}{2}\aalpha_j^{\frac{2}{3}\ell} r^{n-1}\label{defined}
\end{align}

for $r$ small enough. Thus $\delta_2 \w^j_{n-1}$ is defined at $\tilde{y}$, since the latter is in the domain of definition of $\w^j_{n-1}$. It follows that $\mathcal{G}(u) : B_{n-1}^{j,\delta} \rightarrow \HH_{-n+1}$. Moreover
\begin{align}
\Vert \mathcal{G}(u)(y) \Vert_{-n+1} & \leq 2 \sup_{y \in B_{n-1}^{j,\delta}} \Vert \delta_2 w^j_{n-1} \Vert_{-n+1} \leq 2 \varepsilon\aalpha_j^{\ell} r^{\frac{3}{2}(n-1)},
\end{align}
where we used the bounds \eqref{terzo} and \eqref{H}. Hence $\mathcal{G}: \mathcal{B}^j \rightarrow \mathcal{B}^j$.

To prove that $\mathcal{G}$ is a contraction, we use the estimate \eqref{Cauchy3} for 
\begin{align}\tilde{y}_i(y) = \Gamma H\w^j_{n-1}(0) + \widetilde{H}y + \Gamma u_i(y)
\end{align}
and $u_i \in \mathcal{B}, \: i=1,2$. We get immediately that $\tilde{y}_i \in \HH_{n-1}$ and by inequality \eqref{defined}, $\Vert \tilde{y}_i \Vert \leq \frac{1}{2} \aalpha_j^{\frac{2}{3}\ell}r^{n-1}$. Hence the bounds \eqref{Cauchy3}, \eqref{terzo}, \eqref{H}, together with the relations between the $n$-dependent spaces and their norms, imply
\begin{align}
& \Vert \mathcal{G}(u_1) - \mathcal{G}(u_2) \Vert_{\mathcal{B}^j} = \!\!\!\! \sup_{y \in  B_{n-1}^{j,\delta}} \Vert H\delta_2\w^j_{n-1}(\tilde{y}_1) -  H\delta_2\w^j_{n-1}(\tilde{y}_2) \Vert_{-n+1} \notag \\
& \leq 4\aalpha_j^{-\frac{2}{3}\ell}r^{-n+1} \sup_{y \in B_{n-1}^{j} } \Vert\delta_2\w^j_{n-1}(y)\Vert_{-n+1} \sup_{y \in  B_{n-1}^{j,\delta}} \Vert \tilde{y}_1 - \tilde{y}_2\Vert_{-n+1} \notag \\
& \leq 4\aalpha_j^{\frac{1}{3}\ell}\varepsilon r^{\frac{1}{2}(n-1)} \sup_{y \in  B_{n-1}^{j,\delta}} \Vert \tilde{y}_1 - \tilde{y}_2\Vert_{-n+1} \notag \\
& \leq 4\aalpha_j^{\frac{1}{3}\ell} \varepsilon r^{\frac{1}{2}(n-1)} C\eta^{-2n} \sup_{y \in  B_{n-1}^{j,\delta}} \Vert u_1(y) - u_2(y) \Vert_{-n+1} \notag \\
& \leq \frac{1}{2} \Vert u_1(y) - u_2(y) \Vert_{\mathcal{B}^j}
\end{align}
for $r$ and $\varepsilon$ small enough, proving the contractive property of $\mathcal{G}$ on $\mathcal{B}^j$. Hence the existence of the fixed point $u \in \mathcal{B}^j$ of $\mathcal{G}$ solving the equation \eqref{u} and providing $\w^j_n : B_{n-1}^{j,\delta} \rightarrow \HH_{-n+1}$ given by \eqref{decomposition}. Using the natural embeddings we may consider $B_n^j$ a subset of $B_{n-1}^{j,\delta}$, and $\w_n^j$ may be regarded as an element of the space $\mathcal{A}_n^j$. Note also that, since $\tilde y = y + \Gamma \w_n^j(y)$ (see \eqref{u}), the inequality \eqref{defined} can be rewritten as 
\begin{align}
\Vert y + \Gamma \w_n^j(y) \Vert \leq \frac{1}{2}\aalpha_j^{\frac{2}{3}\ell}r^{n-1} \quad \textrm{for} \quad y \in B_n^{j} \label{defined2}
\end{align}  
which implies that $y + \Gamma \w_n^j(y) \in B_{n-1}^j$ for such $y$.
\end{subsection}


\begin{subsection}{Proof of (b)}
In view of the decomposition \eqref{decomposition}, we write
\begin{align}
\w^j_n(y) = \w^j_n(0) + D\w^j_n(0)y + \delta_2 \w^j_n(y), \label{may}
\end{align}
where
\begin{align}
& \w^j_n(0;q) =   H\w_{n-1}^j(0;q) + u(0;q) \notag \\
& D\w^j_n(0) = HD\w^j_{n-1}(0) + Du(0) \notag \\
& \delta_2\w^j_n(y) = \delta_2 u(y) \label{pieces}
\end{align}

Let us first iterate the bound \eqref{primo}. Note that, with the projection $P$ defined at page \pageref{outtheconstants}
\begin{align}
P\w^j_n(0;q) = PHP\w^j_{n-1}(0;q) +  Pu(0;q) \label{lalala}
\end{align}
since $H = HP$. Since $u \in \mathcal{B}^j$ (See definition \eqref{ball}), we have for $0<|\omega \cdot q| \leq \eta^{n-1}$
\begin{align}
| u(0;q) | \leq \Vert u(0) \Vert_{-n+1} \leq 2\varepsilon \aalpha_j^{\ell}r^{\frac{3}{2}(n-1)}. \label{toless}
\end{align}
and Eq. \eqref{lalala}, using the estimate \eqref{primo}, yields 
\begin{align}
|\w^j_n(0;q)|  \leq \left(2^{n}-1\right) \varepsilon \frac{\aalpha_j^{\frac{2}{3}\ell}}{|q|^{\frac{\ell}{3}}} + | u(0;q) |; \label{less}
\end{align}
we omitted here the technical details of the estimate of $PHP\w^j_{n-1}(0;q)$, which is obtained by expanding $H$ in a Neumann series; such details are carried out at p. \pageref{riteration} in the estimate of the quantity \eqref{riteration}. At this point the inequality \eqref{less}, in view of \eqref{toless}, seems less than what we need to iterate \eqref{primo}, but in fact it is much more, as we need a bound only for $|\textrm{Im} \, \beta| \leq \aalpha_{(j;n)}$. For such $\beta$, using the estimate \eqref{toless} we get for $0 < |\omega \cdot q| \leq \eta^{n-1}$
\begin{align}
|u_{\beta}(0;q)| e^{(\bar \aalpha_{j,n-1}- \bar \aalpha_{j,n})|q|} = |u_{\beta'}(0;q)| \leq 2\varepsilon \aalpha_j^{\ell}r^{\frac{3}{2}(n-1)} \label{shrink}
\end{align}
where 
\begin{align}
\beta' = \beta - i \frac{(\bar \aalpha_{j,n-1}- \bar \aalpha_{j,n})}{|q|}q \quad \textrm{so that} \quad |\textrm{Im} \, \beta'| \leq \bar \aalpha_{j,n-1}. 
\end{align}
From the definition \eqref{sequence} we can write $\aalpha_{j,n-1}- \aalpha_{j,n} = \frac{\aalpha_j}{2n(n+1)}$. It follows from \eqref{shrink} that for $0 < |\omega \cdot q| \leq \eta^{n-1}$
\begin{align}
& |u_{\beta}(0;q)| \leq 2\varepsilon \aalpha_j^{\ell}r^{\frac{3}{2}(n-1)} e^{(\frac{\aalpha_j}{2n(n+1)})|q|} \notag \\
& \leq 2\varepsilon \frac{\aalpha_j^{\frac{2}{3}\ell}}{|q|^{\frac{\ell}{3}}}r^{\frac{3}{2}(n-1)} \left[2n(n+1)\right]^{\frac{\ell}{3}} \frac{\ell!}{6}\notag \\
& \leq \varepsilon \frac{\aalpha_j^{\frac{2}{3}\ell}}{|q|^{\frac{\ell}{3}}} \label{sec}
\end{align}
for $r$ small enough. Now, combining \eqref{less} and \eqref{sec} we get the desired bound:
\begin{align}
|\w^j_n(0;q)| \leq \left( 2^{n+1}-1\right) \varepsilon \frac{\aalpha_j^{\frac{2}{3}\ell}}{|q|^{\frac{\ell}{3}}} \quad \text{for} \quad |\omega \cdot q| \leq \eta^{n-1} 
\end{align}

We can now iterate \eqref{terzo} for $\delta_2\w^j_n(y) = \delta_2 u(y)$ (See \eqref{pieces}). We already proved that for $\Vert y \Vert_{n-1} \leq \aalpha_j^{\frac{2}{3}\ell}r^{n-\delta}$ we have $\Vert u(y) \Vert_{-n+1} \leq 2\varepsilon \aalpha_j^{\ell}r^{\frac{3}{2}(n-1)}$ (see \eqref{ball}). We can apply the estimate \eqref{Cauchy2} with $k=2$ and $\gamma = r^{\delta}$, so that for$\Vert y \Vert_n \leq \aalpha_j^{\frac{2}{3}\ell}r^n$ we get 
\begin{align}
\Vert \delta_2\w^j_{-n}(y)\Vert_n & \leq \sup_{\Vert y \Vert_{n-1} \leq \aalpha_j^{\frac{2}{3}\ell}r^n} \Vert \delta_2 u(y) \Vert_{-n+1} \notag \\
& \leq \frac{r^{2\delta}}{1 - r^{\delta}} \sup_{\Vert y \Vert_{n-1} \leq \aalpha_j^{\frac{2}{3}\ell}r^{n-\delta}} \Vert u(y) \Vert_{-n+1} \notag \\
& \leq \frac{r^{2\delta - \frac{3}{2}}}{1 - r^{\delta}} 2\varepsilon \aalpha_j^{\ell}r^{\frac{3}{2}n}.
\end{align}
Taking $\delta > \frac{3}{4}$ and $r$ small enough, we infer that $\Vert \delta_2\w^j_n(y)\Vert_{-n} \leq \varepsilon \aalpha_j^{\ell}r^{\frac{3}{2}n}$, which concludes the inductive proof of (b).

\end{subsection}


\begin{subsection}{Proof of (c)}

This is the part where the identities introduced in section \ref{resonances} are needed.
We will make use of the maps $\pi_{n\beta}: B(r_j^n) \subset \HH \rightarrow \LL(\HH;\HH)$, constructed for $|\lambda| \leq \lambda_n$. In view of the embeddings \eqref{embeddings} such maps can be viewed as
\begin{align}
\pi^j_{n\beta}: B_n^j \subset \HH_n \rightarrow \LL(\HH_n;\HH_{-n}).
\end{align}
We shall show that they can be extended to $|\lambda| \leq \lambda_0$, and the bound \eqref{secondo} will be proven by
\begin{lemma} \label{final}
Denote by $D_n$ the disk $\{\kappa \in \C | |\kappa| \leq \eta^n  \}$ and splitting $\pi^j_{n\beta}(\kappa;0)$ into its diagonal and off diagonal parts 
\begin{align}
\pi^j_{n\beta}(\kappa ; 0) = \sigma^j_{n\beta}(\kappa) + \rho^j_{n\beta}(\kappa),
\end{align}
where $\sigma^j_{n\beta}(\kappa;q,q') = \pi^j_{n\beta}(\kappa ; 0; q,q')\delta_{q,q'}$.
The maps 
\begin{align}
\pi^j_{n\beta} : D_n \times B_n^j \rightarrow \LL(\HH_n;\HH_{-n})
\end{align}
extend analytically to $|\lambda| \leq \lambda_0$, their extensions will still be smooth interpolations of the kernel of $t_p Dw_{n\beta}(y)$, i.e. 
\begin{align}
t_p Dw^j_{n\beta}(y) = \pi^j_{n\beta}(\omega \cdot p;y) \quad \textrm{and} \quad t_p\pi^j_{n\beta}(\kappa;y)= \pi^j_{n\beta}(\kappa +\omega \cdot p;y) \label{interpol}
\end{align} 
 they will depend analytically on $\beta$ and $y$ and belong to $\mathcal{C}^{\infty}(D_n)$. For $i=0,1,2$, they obey the bounds
\begin{align}
& \Vert \partial^i_{\kappa}\delta_1 \pi^j_{n\beta}(\kappa;y) \Vert_{\LL(\HH_n;\HH_{-n})} \leq \varepsilon \aalpha_j^{\frac{\ell}{3}}r^{\frac{1}{2+i}n} \label{piprimo} \\
& \Vert \partial^i_{\kappa}\sigma^j_{n\beta}(\kappa) \Vert_{\LL(\HH_n;\HH_{-n})} \leq  \varepsilon \eta^{(2-i)n} \label{pisecondo} \\
& |\partial^i_{\kappa} \rho^j_{n\beta}(\kappa;q,q') | \leq  \varepsilon \frac{1}{|q-q'|^{\frac{\ell}{3}}}  , \label{piterzo}
\end{align} 
where $\delta_1 \pi^j_{n\beta}(\kappa;y) \equiv \pi^j_{n\beta}(\kappa;y) - \pi^j_{n\beta}(\kappa;0)$.
\end{lemma}

\begin{remark}
By using the diophantine condition as we did at p. \pageref{discussion} in order to get \eqref{oldstyle}, we see that the bound  \eqref{piterzo} implies
\begin{align}
\Vert \partial^i_{\kappa} \rho^j_{n\beta}(\kappa) \Vert_{\LL(\HH_n;\HH_{-n})} \leq  \varepsilon r^{\frac{n}{2}}  , \label{piterzooldstyle}
\end{align}
for $\ell$ large enough.
\end{remark}

Taking $p=0$ in \eqref{interpol} and combining Eqs. \eqref{piprimo}, \eqref{pisecondo} and \eqref{piterzo} we obtain \eqref{secondo}, so we are only left with
\begin{proof}{(Of Lemma \ref{final})}
Differentiating \eqref{iterationt} with respect to $y$ we get
\begin{align}
Dw^j_n(y) = \left(1-Dw^j_{n-1}(\tilde y)\Gamma_{n-1} \right)^{-1}Dw^j_{n-1}(\tilde y) \label{Derivative}
\end{align}
where $\tilde y = y + \Gamma_{n-1}\w_{n\beta}^j(y)$. The right hand side is well defined for $y \in B_{n-1}^{j,\delta} \subset \HH_{n-1}$, in fact by inequality \eqref{defined2}, $\tilde y \in B_{n-1}^j$ for such $y$'s. Lemma \ref{cutoff} and the inductive hypotheses \eqref{secondo} imply that 
\begin{align}
\Vert D\w_n^j(\tilde y)\Gamma_{n-1} \Vert_{\mathcal{L}(H_{-n+1};H_{-n+1})} \leq C\varepsilon \label{bye}
\end{align}

Using the relation \eqref{pirecursive} we define
\begin{align}
\pi_{n\beta}^j(\kappa;y) = \left[1-\pi^j_{(n-1)\beta}(\kappa;\tilde y)\Gamma_{n-1}(\kappa) \right]^{-1} \pi_{(n-1)\beta}^j(\kappa;\tilde y). \label{recursivepi}
\end{align}
The relations \eqref{interpol} follow by simply applying $t_p$ to \eqref{Derivative} and \eqref{recursivepi}.
By the inductive hypotheses, for $\kappa \in D_{n-1}$ and $y \in B_n^{j.\delta}$, $\pi_{n\beta}^j(\kappa;y) \in\LL(\HH_n;\HH_{-n})$ and it is an analytic function of its arguments. Hence, by induction, it coincides for $|\lambda| \leq \lambda_n$ with the maps $\pi_{n\beta}$ constructed in section \ref{resonances}.
Note that
\begin{align}
\pi_{n\beta}^j(\kappa;0) = \left[1-\pi^j_{(n-1)\beta}(\kappa;\tilde 0)\Gamma_{n-1}(\kappa) \right]^{-1} \pi_{(n-1)\beta}^j(\kappa;\tilde 0),
\end{align}    
where $\tilde 0 = \Gamma \w^j_{n\beta}(0)$.

To get an a priori bound from \eqref{recursivepi}, we formulate an easy Lemma
\begin{lemma} \label{Accalemma}
Let  $H_n^j(\kappa, y) \equiv \left[1-\pi^j_{(n-1)\beta}(\kappa; y)\Gamma_{n-1}(\kappa) \right]^{-1}$. For $y \in B_{n-1}^j$ and all $m \leq n$
\begin{align}
\Vert \partial_{\kappa}^i H_m^j(\kappa, y) \Vert_{\mathcal{L}(\HH_{n-1}, \HH_{-n+1})} \leq 2\eta^{-i(m-1)} \quad \textrm{for} \quad i=0,1,2 \label{Acca} 
\end{align}
\end{lemma}

\begin{proof}
For $i=0$ \eqref{bye} implies trivially that $\Vert H_m^j(\kappa, y) \Vert_{\mathcal{L}(\HH_{n-1}, \HH_{-n+1})} \leq 2$. For $i=1$ we have
\begin{align} 
& \Vert \partial_{\kappa} H_m^j(\kappa, y) \Vert_{\mathcal{L}(\HH_{n-1}, \HH_{-n+1})} = \notag \\
& = \Vert H_m^j(\kappa, y)\partial_{\kappa} \!\!\left( \pi^j_{(m-1)\beta}(\kappa; y)\Gamma_{m-1}(\kappa) \right) \!\! H_m^j(\kappa, y) \Vert_{\mathcal{L}(\HH_{n-1}, \HH_{-n+1})} \notag \\
& \leq 2\eta^{-(m-1)}.
\end{align}
In the same fashion one gets 
\begin{align} 
& \Vert \partial^2_{\kappa} H_m^j(\kappa, y) \Vert_{\mathcal{L}(\HH_{n-1}, \HH_{-n+1})} \leq 2\eta^{-2(m-1)}
\end{align}
\end{proof}

From the latter Lemma,  \eqref{recursivepi} and the inductive hypotheses we get the a priori bound
\begin{align}
\Vert \partial^i_{\kappa}\pi^j_{n\beta}(\kappa;y) \Vert_{\LL(\HH_{n-1};\HH_{-n+1})} \leq C \varepsilon  \eta^{(2-i)(n-1)}. \label{apriori}
\end{align}

To prove \eqref{piprimo} we note the identity
\begin{align}
H_n(\kappa; \tilde y)\pi^j_{(n-1)\beta}(\kappa, \tilde y) & = \pi^j_{(n-1)\beta}(\kappa, \tilde y)\left[ 1-\Gamma_{n-1}\pi^j_{(n-1)\beta}(\kappa;\tilde y)  \right]^{-1} \notag \\
& \equiv  \pi^j_{(n-1)\beta}(\kappa, \tilde y) \widetilde H_n^j (\kappa; \tilde y),
\end{align}
 which, for $y \in B_{n-1}^{j,\delta}$ yields
\begin{align}
& \delta_1 \pi^j_{n\beta} (\kappa;y) = \pi^j_{n\beta} (\kappa;y)-\pi^j_{n\beta} (\kappa;0) \notag \\
& = H_n^j (\kappa; \tilde y) \pi^j_{(n-1)\beta} (\kappa;\tilde y) - \pi^j_{(n-1)\beta} (\kappa;\tilde 0)\widetilde H_n^j (\kappa; \tilde 0) \notag \\
& = H_n^j (\kappa; \tilde y) \!\left[ \pi^j_{(n-1)\beta} (\kappa;\tilde y)(\widetilde H_n^j)^{-1} (\kappa; \tilde 0) \!-\! (H_n^j)^{-1} (\kappa; \tilde y)\pi^j_{(n-1)\beta} (\kappa;\tilde 0) \right]\!\widetilde H_n^j (\kappa; \tilde 0) \notag \\
& = H_n^j (\kappa; \tilde y) \Big[ \pi^j_{(n-1)\beta} (\kappa;\tilde y) \left(1-\Gamma_{n-1}(\kappa)\pi^j_{(n-1)\beta}(\kappa; \tilde0)\right) \notag \\
& \quad - \left(1-\pi^j_{(n-1)\beta}(\kappa; \tilde y)\Gamma_{n-1}(\kappa)\right)\pi_{(n-1)\beta}^j(\kappa; \tilde 0)\Big]\widetilde H_n^j (\kappa; \tilde 0) \notag \\
&  = H_n^j (\kappa; \tilde y) \left[  \pi^j_{(n-1)\beta} (\kappa;\tilde y) - \pi^j_{(n-1)\beta} (\kappa;\tilde 0) \right] \widetilde H_n^j (\kappa; \tilde 0)\notag \\
&  = H_n^j (\kappa; \tilde y) \left[  \delta_1\pi^j_{(n-1)\beta} (\kappa;\tilde y) - \delta_1\pi^j_{(n-1)\beta} (\kappa;\tilde 0) \right]\widetilde H_n^j (\kappa; \tilde 0) \label{manipol}.
\end{align}
From Lemma \ref{Accalemma} with $i=0$ the inductive hypotheses and \eqref{manipol} we get the a priori bound for $y \in B_{n-1}^{j,\delta}$
\begin{align}
\Vert \delta_1 \pi^j_{n\beta} (\kappa;y) \Vert_{\LL(\HH_{n};\HH_{-n})} \leq \Vert \delta_1 \pi^j_{n\beta} (\kappa;y) \Vert_{\LL(\HH_{n-1};\HH_{-n+1})} \leq 8\varepsilon \aalpha_j^{\frac{\ell}{3}}r^{\frac{n-1}{2}}. 
\end{align}
To get \eqref{piprimo} with $i=0$, we restrict to $y \in B_n^j$ and using \eqref{Cauchy2} we extract
\begin{align}
& \Vert \delta_1 \pi^j_{n\beta} (\kappa;y) \Vert_{\LL(\HH_{n};\HH_{-n})} = \Vert \delta_1 \delta_1 \pi^j_{n\beta} (\kappa;y)\Vert_{\LL(\HH_{n};\HH_{-n})} \notag \\
& \leq \sup_{y \in B_n^j} \Vert \delta_1 \delta_1 \pi^j_{n\beta} (\kappa;y)\Vert_{\LL(\HH_{n};\HH_{-n})} \notag \\
& \leq \frac{r^{\delta}}{1-r^{\delta}} \sup_{y \in B_{n-1}^{j,\delta}} \Vert \delta_1 \pi^j_{n\beta} (\kappa;y)\Vert_{\LL(\HH_{n};\HH_{-n})} \notag \\
& \leq \frac{r^{\delta}}{1-r^{\delta}} 8\varepsilon \aalpha_j^{\frac{\ell}{3}}r^{\frac{n-1}{2}} \leq \varepsilon \aalpha_j^{\frac{\ell}{3}}r^{\frac{n}{2}}.
\end{align}
To get \eqref{piprimo} with $i=1$ we first obtain another a priori bound for $y \in B_{n-1}^{j,\delta}$ by differentiating \eqref{manipol} with respect to $\kappa$ and using \eqref{apriori} and the inductive hypotheses:
\begin{align}
& \Vert \partial_{\kappa} \delta_1 \pi^j_{n\beta} (\kappa;y) \Vert_{\LL(\HH_{n};\HH_{-n})} = \notag \\
& = \Vert  \partial_{\kappa}H_n^j (\kappa; \tilde y) \left[  \delta_1\pi^j_{(n-1)\beta} (\kappa;\tilde y) - \delta_1\pi^j_{(n-1)\beta} (\kappa;\tilde 0) \right]\widetilde H_n^j (\kappa; \tilde 0) \notag \\
& \quad +  H_n^j (\kappa; \tilde y) \partial_{\kappa}\left[  \delta_1\pi^j_{(n-1)\beta} (\kappa;\tilde y) - \delta_1\pi^j_{(n-1)\beta} (\kappa;\tilde 0) \right]\widetilde H_n^j (\kappa; \tilde 0) \notag \\
& \quad +  H_n^j (\kappa; \tilde y) \left[  \delta_1\pi^j_{(n-1)\beta} (\kappa;\tilde y) - \delta_1\pi^j_{(n-1)\beta} (\kappa;\tilde 0) \right]\partial_{\kappa}\widetilde H_n^j (\kappa; \tilde 0) \Vert_{\LL(\HH_{n};\HH_{-n})} \notag \\
& \leq 4\eta^{-n}\varepsilon \aalpha_j^{\frac{\ell}{3}} r^{\frac{n-1}{2}} + 4\varepsilon \aalpha_j^{\frac{\ell}{3}}r^{\frac{n-1}{3}}\eta^{-n}  + 4\eta^{-n}\varepsilon \aalpha_j^{\frac{\ell}{3}} r^{\frac{n-1}{2}}\notag \\
& \leq 12 \varepsilon \aalpha_j^{\frac{\ell}{3}} r^{\frac{n-1}{3}},
\end{align}
then we consider again the ball $B_n^{j}$ to squeeze the correct estimate out:
\begin{align}
\Vert \partial_{\kappa} \delta_1 \pi^j_{n\beta} (\kappa;y) \Vert_{\LL(\HH_{n};\HH_{-n})} \leq \frac{r^{\delta}}{1-r^{\delta}}12 \varepsilon \aalpha_j^{\frac{2}{3}\ell} r^{\frac{n-1}{3}} \leq \varepsilon \aalpha_j^{\frac{2}{3}\ell} r^{\frac{n}{3}}.
\end{align}
The same procedure (establish an a priori bound, then restrict the domain of $y$'s) yields \eqref{piprimo} with $i=2$.

Leaving the more difficult bound \eqref{pisecondo} for last, we can now iterate \eqref{piterzo}. In order to do that inductively,  we write
\begin{align}
\rho^j_{n\beta}(\kappa) = \underbrace{\left[ 1- \pi(0;\kappa)_{(n-1)\beta}^j\Gamma_{n-1}(\kappa)\right]^{-1}\pi^j_{(n-1)\beta}(0;\kappa)}_{\Upsilon_n^j(\kappa)} + R_n^j(\kappa) \label{riteration}
\end{align}
where
\begin{align}
R_n^j(\kappa; q,q') \equiv \left( \frac{1}{1-(\pi(0) + \delta\pi)\Gamma}\delta\pi \Gamma\frac{1}{1-\pi(0)\Gamma}\pi(0) + \frac{1}{1-(\sigma + \delta\pi)\Gamma} \delta \pi \right) \label{roest}
\end{align}
with $\pi = \pi^j_{(n-1)\beta}(\kappa)$ e $\delta \pi = \delta_1 \pi^j_{(n-1)\beta}(\kappa; \tilde 0)$. Using the inductive hypotheses it is not hard to show that
\begin{align}
\Vert \partial_{\kappa}^i R_n^j(\kappa) \Vert_{\LL(\HH_{n};\HH_{-n})} \leq \varepsilon \aalpha_j^{\frac{\ell}{3}} \label{pre}.
\end{align}
In order to estimate the first term in \eqref{riteration} we notice that it can be written as

\begin{align}
\Upsilon_n^j(\kappa; q,q')= \sum_{k=0}^{\infty} \sum_{q_1, \ldots, q_k}  \pi(0; q,q_1)\Gamma(q_1) \cdots \pi(0; q,q_n)\Gamma_{n-1}(q_n)\pi(0; q_n,q') 
\end{align}
where, again, $\pi = \pi^j_{(n-1)\beta}(\kappa)$. The $k$-th term in the series reads (leaving the sums over repeated $q_j$'s understood) \footnote{To be \textit{very} exact and consistent with the expression if $k$ is not even, we should take the sum over $0\leq i_1 \leq i_2 \cdots \leq i_{k+1}=k$, and perform some formal changes in a couple of subindices; we hope the reader will forgive us.}
\begin{align}
& \sum_{0\leq i_1 \leq i_2 \cdots \leq i_k=k} \left[\sigma(q)\Gamma(q)\right]^{i_1}\rho(q, q_{i_1})\Gamma(q_{i_1}) \cdots  \rho(q_{i_2-1},q_{i_2})\Gamma(q_{i_2}) \notag \\
& \quad  \left[\sigma(q_{i_2})\Gamma(q_{i_2})\right]^{i_3-i_2} \rho(q_{i_2},q_{i_3} )\Gamma(q_{i_3}) \cdots  \rho(q_{i_4-1},q_{i_4})\Gamma(q_{i_4}) \notag \\
& \quad \cdots \left[\sigma(q_{i_4})\Gamma(q_{i_4})\right]^{i_{k-1}-i_{k-2}} \rho(q_{i_{k-2}},q_{i_{k-1}} )\Gamma(q_{i_{k-1}}) \cdots \rho(q_{k},q'),
\end{align}
Using the inductive hypothesis again, and the diophantine condition \eqref{Diophantine}, we get 
\begin{align}
    & \Upsilon_n^j(\kappa; q,q') \leq \varepsilon \sum_{k=0}^{\infty} \varepsilon^k\sum_{j=1}^k  \sum_{|\omega \cdot q_i| \leq \eta^{n-1}}\frac{\eta^{-2n}}{|q-q_1|^{\frac{\ell}{3}}}\frac{\eta^{-2n}}{|q_1-q_2|^{\frac{\ell}{3}}} \cdots \frac{\eta^{-2n}}{|q_j-q'|^{\frac{\ell}{3}}} \\
    &  \overset{(*)}{\leq}  \varepsilon \sum_{k=0}^{\infty} \varepsilon^k \sum_{j=1}^k \left[ \eta^{-2n} 2^{\ell}\left( 2\gamma^{-1}\eta^{n-1} \right)^{ \frac{\ell}{\nu}\left(\frac{\ell}{3}-d \right)}\right]^j \frac{1}{|q-q'|^{\frac{\ell}{3}}} \notag \\
    & \leq \frac{1}{2} \varepsilon  \frac{1}{|q-q'|^{\frac{\ell}{3}}}  \label{pre2}
    \end{align}
    for $\ell$ large enough and $\varepsilon$ small enough. To obtain (*) we repeatedly used the estimate
        \begin{align}
    \sum_{|\omega \cdot p| \leq \eta^{n-1}} \frac{1}{|q-p|^{\frac{\ell}{3}}} \frac{1}{|p-q'|^{\frac{\ell}{3}}} \leq \frac{2^{\ell}\left( 2\gamma^{-1}\eta^{n-1} \right)^{ \frac{\ell}{\nu}\left(\frac{\ell}{3}-d \right)}}{|q-q'|^{\frac{\ell}{3}}}     \end{align}
     for all $|\omega \cdot q|,\,|\omega \cdot q'| \leq \eta^{n-1} \quad  \text{and} \quad q \neq q'$, which is obtained by using the diophantine condition as in \eqref{oldstyle} and Minkowski inequality for the $\ell^p$ spaces: $\Vert f + g \Vert_{p} \leq \Vert f \Vert_{p} + \Vert g \Vert_{p}$. Now combining \eqref{pre} and \eqref{pre2} we get
\begin{align}
|\rho_{n \beta} (\kappa; q,q')| \leq \varepsilon \aalpha_j^{\frac{\ell}{3}}  + \frac{1}{2} \varepsilon  \frac{1}{|q-q'|^{\frac{\ell}{3}}} \quad \text{for} \quad |\omega \cdot q| , \, |\omega \cdot q'| \leq \eta^{n-1}. \label{notyet}
\end{align}
Reasoning exactly in the same way we did at p. \pageref{toless}, we notice that the last bound holds for all $|\text{Im}\, \beta | \leq \bar \alpha_{(j;n-1)}$, hence we can shift $\beta$, and making use of the diophantine property of $\omega$ (Cf. p.\pageref{shrink}) we get  for $|\text{Im}\, \beta | \leq \bar \alpha_{(j;n)}$
\begin{align}
|\rho_{n \beta} (\kappa; q,q')| \leq  \varepsilon  \frac{1}{|q-q'|^{\frac{\ell}{3}}} \quad \text{for} \quad |\omega \cdot q| , \, |\omega \cdot q'| \leq \eta^{n-1}. \label{yet},
\end{align} 
that is, \eqref{piterzo} for $i=0$. Without any difference one obtains \eqref{pre2} for $\partial_{\kappa}\rho$ and $\partial^2_{\kappa}\rho$, which combined with \eqref{pre} and the diophantine condition on $\omega$ (see \eqref{notyet}-\eqref{yet}) yields \eqref{piterzo} for $i=1,2$.

To prove \eqref{pisecondo} we need to establish a Lemma that will follow from the discussion of section \ref{WWW} as a consequence of the Ward identity \eqref{Ward2}(the indices $j$ are omitted and the upper indeces stand for the components):
\begin{lemma}
Assume the inductive hypotheses \eqref{xdecay}, $|\textrm{Im}\beta| \leq \alpha_{(j;n)}$ and $\varepsilon$ sufficiently small, then the following inequalities hold
\begin{align}
&  | \sigma_{n\beta}(0;0) | \leq \varepsilon r^{\frac{n}{2}}, \label{consequence1} \\
&  | \partial_{\kappa}\sigma_{n\beta}(0;0) | \leq \varepsilon  \eta^{2n}. \label{consequence2}
\end{align}
\end{lemma}
\begin{proof}
Using Eq.\eqref{Ward2} evaluated at $q=0$, we get
\begin{align}
& \sigma_{n}^{\gamma,\alpha}(\kappa;0)\Big|_{\kappa = 0} =\pi_{n}^{\gamma,\alpha}(\kappa;y;0,0)\Big|_{\substack{\kappa = 0\\ y=0}} = Dw_{n}^{\gamma, \alpha}(y;0,0)\Big|_{y=0} \notag \\
& =  - \sum_{q\in \Z^d}iq^{\gamma} \bar{\chi}_n(\omega \cdot q)\x^{\beta}(q)\rho_n^{\beta, \alpha}(0;-q,0) ,
\end{align}
so \eqref{consequence1} follows from the decay of the coefficients $\x(q)$ and from \eqref{piterzooldstyle}
\begin{align}
| \sigma_{n\beta}(0;0) | \leq \Vert \rho_n (0) \Vert_{\LL(\HH_{n};\HH_{-n})} \leq  \varepsilon  r^{\frac{n}{2}}.
\end{align}

Using \eqref{piderivative} we get
\begin{align}
 \partial_{\kappa} \sigma_n (0;0)^{\alpha,\gamma} & = \sum_{q} \pi_{n}^{\alpha, \delta}(0 ; 0 ; q,0) \partial_{\kappa}\gamma_{<n}(\omega \cdot q)\pi_{n}^{\gamma, \delta}(0;0;-q;0) \notag \\
& = \sum_{q} D\w_{n}^{\alpha, \delta}(0 ; q,0) \partial_{\kappa}\gamma_{<n}(\omega \cdot q)D\w_n^{\gamma, \delta}(0;-q;0),
\end{align}
using \eqref{Ward2} the latter takes the form
\begin{align}
\partial_{\kappa} \sigma_n (0;0)= \mathcal{Z}_n + \mathcal{Q}_n,
\end{align}
where
\begin{align}
\mathcal{Z}_n^{\alpha,\gamma} & = \!-\!\sum_{q}q^{\alpha}q^{\gamma} \!\left(w_0^{\delta}(\x + \Gamma_{<n}\w_n(0);-q)\right) \times \notag \\
& \quad \times \partial_{\kappa}\gamma_{<n}(\omega \cdot q)\! \left(w_0^{\delta}(\x + \Gamma_{<n}\w_n(0);-q)\right)   \label{Z} \\
\text{and} \notag \\
 \mathcal{Q}_n^{\alpha,\gamma} & =  \!\!\! \sum_{q, q', q''}iq'^{\alpha} \bar{\chi}_n(\omega \cdot q')\x^{\beta}(q')\pi_n^{\beta, \delta}(0;0;q',q) \partial_{\kappa} \gamma_{<n}(\omega \cdot q) \cdot  \\
& \quad  \cdot iq''^{\gamma} \bar{\chi}_n(\omega \cdot q'')\x^{\beta'}(q'')\pi_n^{\beta',\delta}(0;0;q'',q). \notag
\end{align}
The expression summed in the right hand side of \eqref{Z} is odd in $q$, hence $\mathcal{Z}_n$ vanishes, so, using Lemma \ref{cutoff} and \eqref{apriori}, we have 
\begin{align}
& | \partial_{\kappa} \sigma_n (0;0) | = | \mathcal{Q}_n | \leq \Vert \pi_n \Vert_{\LL(\HH_{n};\HH_{-n})} \Vert \Vert \partial_{\kappa}\Gamma_{<n}(\kappa) \Vert \Vert \pi_n \Vert_{\LL(\HH_{n};\HH_{-n})} \notag \\
& \leq C\varepsilon^2 \eta^{2n-2} \leq \varepsilon \eta^{2n}
\end{align}
for $\varepsilon$ small enough.
\end{proof}

Using \eqref{recursivepi} we write 
\begin{align}
\sigma^j_{n\beta}(\kappa) = \underbrace{\left[ 1- \sigma_{(n-1)\beta}(\kappa)\Gamma_{n-1}(\kappa)\right]^{-1}}_{K_{n\beta}^j(\kappa)}\sigma_{(n-1)\beta}(\kappa) + S_n^j(\kappa) \label{siteration}
\end{align}
where
\begin{align}
S_n^j(\kappa) \equiv \textrm{diag}\left( \frac{1}{1-(\sigma + \mathcal{R})\Gamma}\mathcal{R}\Gamma\frac{1}{1-\sigma\Gamma}\sigma + \frac{1}{1-(\sigma + \mathcal{R})\Gamma} \mathcal{R} \right) \label{rest}
\end{align}
with $\sigma = \sigma^j_{(n-1)\beta}(\kappa)$ e $\mathcal{R}= \rho^j_{(n-1)\beta}(\kappa) +  \delta_1 \pi^j_{(n-1)\beta}(\kappa; \tilde 0)$. Using the inductive hypotheses it is not difficult to show that
\begin{align}
\Vert \partial_{\kappa}^i S_n^j(\kappa) \Vert_{\LL(\HH_{n};\HH_{-n})} \leq \varepsilon r^{\frac{n}{2}} \label{notdifficult}
\end{align}
as $\mathcal{R}$ appears as a factor in both terms of \eqref{rest}.

We shall now describe a crucial property of $K_n^j(\kappa)$: fixing $n$ and $|\kappa| \leq \eta^n$, we have that $K_m^j (\kappa; q)$ restricted to the set $\{ q \in \Z^d : |\omega \cdot q| \leq \eta^{n-1}\}$, is the identity for all $m \leq n-2$. In fact, for such $\kappa$'s and $q$'s, we have $|\omega \cdot q + \kappa| \leq \eta^{n-2}$. On the other hand $\Gamma_{m}(\kappa)$ is supported on the set $|\omega \cdot q + \kappa| \geq \eta^m$, i.e. whenever $\eta^{n-2} \leq \eta^m$, we have $\Gamma_{m}(\kappa) = 0$. Summarizing for $m \leq n-2$ and $|\kappa| \leq \eta^n$
\begin{align}
K_m^j (\kappa;q) = \left[ 1- \sigma_{(n-1)\beta}(\kappa)\Gamma_{n-1}(\kappa)\right]^{-1}(q) = \textrm{Id}(q), \quad \textrm{for} \quad |\omega \cdot q| \leq \eta^{n-1} . \label{identity}
\end{align}
So, for all $m \leq n-2$ and $|\kappa| \leq \eta^n$ we have
\begin{align}
\sigma^j_{m\beta}(\kappa;q) = \sigma^j_{(m-1)\beta}(\kappa;q) + R_m^j(\kappa;q)  \quad \textrm{for}  \quad |\omega \cdot q| \leq \eta^{n-1} .\label{fundamental}
\end{align}

In view of \eqref{identity} we notice that "on the scale $n$", $\sigma_m$ stays almost constant until $m=n-2$, in fact if we assume $\Vert \partial^2 \sigma^j_{0\beta}(\kappa) \Vert_{\LL(\HH_{n};\HH_{-n})} \leq \frac{1}{16}\varepsilon $ which we can always do, it follows from \eqref{fundamental} and \eqref{notdifficult},
\begin{align}
\Vert \partial_{\kappa}^2 \sigma^j_{(n-2)\beta}(\kappa) \Vert_{\LL(\HH_{n};\HH_{-n})} \leq \varepsilon \left(\frac{1}{16} + \sum_{k=1}^{n-2}r^{\frac{k}{2}}\right) = \frac{1}{8}\varepsilon. \label{almost2}
\end{align}

Now we can prove \eqref{pisecondo}. For $i=0$ we use \eqref{siteration} twice and make use of the fact that for all $m$ $\sigma^j_m(q;\kappa) =\sigma^j_m(0;\tilde \kappa) $ with $\tilde \kappa := \kappa + \omega \cdot q$, so we get
\begin{small}
\begin{align}
& \sigma^j_{n\beta}(\kappa) = K_{n\beta}^j(\kappa)K_{(n-1)\beta}^j(\kappa)\sigma_{(n-2)\beta}(\kappa) + K_{n\beta}^j(\kappa) S_{n-1}^j(\kappa) + S_n^j(\kappa) \notag \\
& = \! K_{n\beta}^j(\kappa)K_{(n-1)\beta}^j(\kappa)\!\! \left(\int_0^{\tilde \kappa} \!\!\! \int_0^{\kappa'} \!\!\!\partial^2\sigma_{(n-2)\beta}(\kappa'';0)d\kappa'' \, d\kappa' + \tilde \kappa   \partial \sigma_{(n-2)\beta}(0;0) + \sigma_{(n-2)\beta}(0;0) \!\!  \right) \notag \\
& \quad + K_{n\beta}^j(\kappa) S_{n-1}^j(\kappa) + S_n^j(\kappa) \label{sibetter}
\end{align}
\end{small}
from which, using Lemma \ref{Accalemma}, \eqref{consequence1}, \eqref{consequence2} and \eqref{notdifficult} we get
\begin{align}
& \Vert \sigma^j_{n\beta}(\kappa) \Vert_{\LL(\HH_{n};\HH_{-n})} \leq \varepsilon \left( \frac{\tilde \kappa^2}{4} + \tilde \kappa \eta^{2(n-2)}  +r^{\frac{n-2}{2}} \right) + 2 \varepsilon r_j^{\frac{n-1}{2}} +  \varepsilon r_j^{\frac{n}{2}} \notag \\
& \leq  \varepsilon \eta^{2n}. \label{doneit}
\end{align}
Differentiating \eqref{sibetter} with respect to $\kappa$ and using Lemma \ref{Accalemma}, \eqref{consequence1}, \eqref{consequence2} and \eqref{notdifficult}, we get
\begin{align}
 & \partial_{\kappa} \sigma^j_{n\beta}(\kappa) = \partial_{\kappa}\left(K_{n\beta}^j(\kappa)K_{(n-1)\beta}^j(\kappa)\right)\sigma_{(n-2)\beta}(\kappa)\notag \\
& \quad + K_{n\beta}^j(\kappa)K_{(n-1)\beta}^j(\kappa) \partial_{\kappa}\sigma_{(n-2)\beta}(\kappa) + \partial_{\kappa}K_{n\beta}^j(\kappa) S_{n-1}^j(\kappa) \notag \\
& \quad + K_{n\beta}^j(\kappa) \partial_{\kappa} S_{n-1}^j(\kappa) + \partial_{\kappa} R_n^j(\kappa),
\end{align}
and proceeding as in \eqref{doneit} we get
\begin{align}
& \Vert \partial_{\kappa} \sigma^j_{n\beta}(\kappa) \Vert_{\LL(\HH_{n};\HH_{-n})} \leq  \varepsilon \eta^n.
\end{align}
In the same way we get obtain the bound
\begin{align}
\Vert \partial_{\kappa}^2 \sigma^j_{n\beta}(\kappa)\Vert_{\LL(\HH_{n};\HH_{-n})} \leq \varepsilon.
\end{align}
which concludes the proofs of Lemma \ref{final}, of \textbf{(c)} (p. \pageref{etichetta})  and, hence, of Proposition \ref{main}.
\end{proof}

\end{subsection}
\end{section}

\begin{section}{Proof of Theorem \ref{KAM}} \label{finalsection}

In this section we shall show that $Y^j_n \equiv F^j_n(0)$ converges to an analytic function $Y^j$ with zero average for $n \rightarrow \infty$, solving \eqref{EquazDelta}. Furthermore $X_j \equiv \sum_{i=0}^j Y^i$ converges to a differentiable function $X$  with zero average for $j \rightarrow \infty$, solving \eqref{Prob}, which proves Theorem \ref{KAM}.

First of all in Proposition \ref{easy} we constructed for $|\lambda| \leq \lambda_n$ the analytic maps $f_{n\beta}^j$ from $B_n^j \subset \HH$ to $\HH$, satisfying the relations \eqref{iterationf} and \eqref{fcumulative} and obeying the bound 
\begin{align}
\sup_{y \in B_n^j} \Vert f^j_{n\beta} \Vert \leq 2\aalpha_j^{\frac{2}{3}\ell}r^n.
\end{align}
They may be also viewed as analytic maps from $B_n^j \subset \HH_n$ to $\HH$. As such they may be analytically extended to $|\lambda| \leq \lambda_0$ for $n \geq n_0$ by iterated use of \eqref{iterationf} if we recall the bound \eqref{defined2}. The new maps are clearly bounded uniformly in $n$ (e.g. by $2\aalpha_j^{\frac{2}{3}\ell}r_j^{n_0}$). Let us prove now the convergence in $\HH$ of $y^j_{n\beta} \equiv f^j_{n\beta}(0)$ obtained this way. The recursion \eqref{iterationf} implies
\begin{align}
y^j_{n\beta} = f^j_{n\beta}(0) =  f^j_{(n-1)\beta}(\Gamma_{n-1}\w^j_{n\beta}(0)) = y^j_{(n-1)\beta} + \delta_1f^j_{(n-1)\beta}(\Gamma_{n-1}\w^j_{n\beta}(0)).
\end{align}      
Using Lemma \ref{cutoff}, the bound \eqref{oldstyle}  and \eqref{Cauchy2} we infer
\begin{align}
&  \Vert y^j_{n\beta} - y^j_{(n-1)\beta} \Vert  = \Vert \delta_1f^j_{(n-1)\beta}(\Gamma_{n-1}\w^j_{n\beta}(0)) \Vert \notag \\
& \leq \sup_{\Vert y \Vert \leq C\eta^{-2n} \varepsilon \aalpha_j^{\frac{2}{3}\ell} r^{2n}} \Vert \delta_1f^j_{(n-1)\beta}(y) \Vert \notag \\
& \leq \frac{C\eta^{-2n}\varepsilon  r^{n+1}}{1-C\eta^{-2n}\varepsilon r^{n+1}}\sup_{\Vert y \Vert \leq \aalpha_j^{\frac{2}{3}\ell}r^{n-1}} \Vert f^j_{(n-1)\beta}(y) \Vert \leq C\eta^{-2n}\varepsilon\aalpha_j^{\frac{2}{3}\ell} r^{n}
\end{align}
The sequence is hence Cauchy, and therefore it converges in $\HH$:
\begin{align}
y^j_{n\beta} \overset{n\rightarrow \infty }{\longrightarrow} y^j_{\beta}
\end{align}
with 
\begin{align}\Vert y^j_{\beta} \Vert \leq C\varepsilon \aalpha_j^{\frac{2}{3}\ell} \label{BernMos} 
\end{align}
uniformly in the strip $|\textrm{Im} \, \beta| \leq \frac{1}{2}\bar{\alpha}_{j}$. This last estimate implies that, pointwise,
\begin{align} 
|y^j(q)| \leq C\varepsilon \aalpha_j^{\frac{2}{3}\ell} e^{-\frac{\bar{\alpha}_j}{2}|q|} \label{BernMospoint}.
\end{align}

For $|\lambda| \leq \lambda_n $, Eqs \eqref{fcumulative} and \eqref{titcumulative} imply that
\begin{align}
y^j_n \equiv f^j_n(0) = \Gamma_{<n}\w_0^j(y^j_n) \quad \textrm{and} \quad \w_0^j(y^j_n) = \w_n^j(0) \label{head}.
\end{align}
From the first Eq. in \eqref{head} we get $y^j_n(q)|_{q=0} = 0$ and from the second one using \eqref{Ward1} and \eqref{oldstyle} it follows
\begin{align}
& \left|\w_0^j(y^j_n; 0)\right|= \left|\w_n^j(0;0)\right| = \left|\sum_{q \neq 0}q \, \bar{\chi}_n(\omega \cdot q)\x(q)\cdot\w_n^j(0;-q)\right| \notag \\
& \leq \Vert P\w^j_n \Vert_{-n} \leq \varepsilon \aalpha_j^{\frac{2}{3}\ell}r^{2n}
\end{align}
By analyticity these relations have to hold also for $|\lambda| \leq \lambda_0$, so we can take the limit for $n \longrightarrow \infty$ in Eqs. \eqref{head} and infer that
\begin{align}
y^j(0) = 0 \, , \quad y^j = G_0\w_0^j(q;y^j) \: \textrm{for} \; q \neq 0 \label{constructed}.
\end{align}
Once we have constructed inductively $y^j(0)$ we set $x_j \equiv y^j + x_{j-1}$; using \eqref{constructed}, the inductive hypotheses on $x_{j-1}$ and \eqref{tildedef} we get $x_j(q)|_{q=0}=0$ and for $q \geq 0$,
\begin{align}
x_j & = y^j + x_{j-1} = G_0\w_0^j(q;y^j) + G_0w_0^{j-1}(q;x_{j-1}) \notag \\
& = G_0w_0^j(q;x_{j-1} + y^j) - G_0w_0^{j-1}(q; x_{j-1}) + G_0w_0^j(q;x_{j-1}) \notag \\
& =  G_0w_0^j(q;x_j) \label{soluzionej}, 
\end{align} 
so $x_j$ solves \eqref{AppSol} for $k=j$. Furthermore, using \eqref{xdecay} and \eqref{BernMospoint} we get
\begin{align}
 |x_j(q)| & \leq  |y^j(q)| + |x_{j-1}(q)| \leq C\varepsilon \aalpha_j^{\frac{2}{3}\ell} e^{-\frac{\bar{\alpha}_j}{2} |q|} + C\varepsilon A_{j} \frac{e^{-\frac{|q|}{4\gamma_j}}}{|q|^{\frac{\ell}{3}}} \notag \\
& \leq  C\varepsilon \ell! \left(  \frac{4\alpha_j^2}{\bar \alpha_j}\right)^{\frac{\ell}{3}}  \frac{e^{-\frac{\bar \alpha_j}{4}|q|}}{|q|^{\frac{\ell}{3}}}+   C\varepsilon \sum_{k=0}^{j-1}  \ell! \left( \frac{4}{M8^{k-5}} \right)^{\frac{\ell}{3}} \frac{e^{-\frac{|q|}{4\gamma_j}}}{|q|^{\frac{\ell}{3}}} \notag \\
& = C\varepsilon \ell! \left( \frac{4}{M8^{j-5}} \right)^{\frac{\ell}{3}} \frac{e^{-\frac{\bar \alpha_j}{4}|q|}}{|q|^{\frac{\ell}{3}}} + C\varepsilon \sum_{k=0}^{j-1}  \ell! \left( \frac{4}{M8^{k-5}} \right)^{\frac{\ell}{3}} \frac{e^{-\frac{|q|}{4\gamma_j}}}{|q|^{\frac{\ell}{3}}} \notag \\
& \leq C\varepsilon A_{j+1} \frac{e^{-\frac{|q|}{4\gamma_{j+1}}}}{|q|^{\frac{\ell}{3}}},
\end{align}
that is \eqref{xdecay} for $x_j$.

If we can show that $x_j$ converges for $j \rightarrow \infty$ to some function $x$, we can take the limit for $j \rightarrow \infty$ on both sides of \eqref{soluzionej} to obtain
\begin{align}
x(0) = 0 \, , \quad x = G_0w_0(q;x) \: \textrm{for} \; q \neq 0 \label{soluzione}
\end{align}
which is the Fourier transformed version of \eqref{Prob}. 
To conclude the proof of Theorem \eqref{KAM} we only have to show that for $j \rightarrow \infty$,  $x_j(q) \rightarrow x(q)$, for all $q \neq 0$, with $\sum_{q \in \Z^d}|q|^s|x(q)| < \infty$ (which implies $X \in \mathcal{C}^s$). In order to do that, we define $u_j := x_j - x_0$ so that   
\begin{align}
\lim_{j\rightarrow \infty} x_j -x_0 =\lim_{j\rightarrow \infty} u_j = \sum_{j=1}^{\infty} u_j -u_{j-1}, 
\end{align}
and using \eqref{BernMospoint} we get, for all $s$
\begin{align}
& \left|\left(u_j(q) -u_{j-1}(q)\right)\right| = \left| y^j(q)\right| \leq  C\varepsilon \aalpha_j^{\frac{2}{3}\ell} e^{-\frac{\bar{\alpha}_j}{2}|q|}\notag \\
& \leq 2^{s+d}8^{3(s+d)} (s+d)!  C\varepsilon \frac{\aalpha_j^{\frac{2}{3}\ell-s-d}}{|q|^{s+d}}\notag \\
& =  C_{s,d} \varepsilon \frac{\aalpha_j^{\frac{2}{3}\ell-s-d}}{|q|^{s+d}}\label{Cauchyfinal},
\end{align}
from the latter bound we get for $s < \frac{2}{3}\ell$, 
\begin{align}
\sum_{q \in \Z^d}|q|^{s} \lim_{j\rightarrow \infty} |u_j(q)|& \leq  \sum_{q \in \Z^d}\sum_{j=1}^{\infty} |q|^s\left|u_j(q) -u_{j-1}(q)\right| \notag \\
& \leq \sum_{q \in \Z^d}\sum_{j=1}^{\infty}    C_{s,d}\varepsilon \frac{\aalpha_j^{\frac{2}{3}\ell-s-d}}{|q|^{d}} < \infty. 
\end{align}
Finally 
\begin{align}
& \sum_{q \in \Z^d}|q|^{s} |x(q)| = \sum_{q \in \Z^d}|q|^{s} \left|\lim_{j\rightarrow \infty} x_j(q)\right| = \sum_{q \in \Z^d}|q|^{s} \left|\lim_{j\rightarrow \infty} u_j(q) + x_0(q) \right| < \infty
\end{align}
which implies that $X \in \mathcal{C}^s$ and proves Theorem \ref{KAM}. 

\hfill $\blacksquare$
\end{section}

\renewcommand{\theequation}{\arabic{section}.\arabic{equation}}
\renewcommand{\thesubsection}{A-\arabic{subsection}}
  \setcounter{equation}{0}  
 \setcounter{section}{10}
 \section*{APPENDIX}  

\subsection{Proof of Lemma \ref{aprioridecay}}\label{Appriori}

First of all we notice that, if $|\text{Im}\, \xi|\leq \frac{1}{4\gamma_j}$ then $|\text{Im}\, (\xi + \X(\xi))|\leq \frac{1}{2\gamma_j}$, in fact
\begin{align}
&\left| \text{Im} \X(\xi) \right| = \left| \text{Im}\left(\X(\xi) - \X(\text{Re} \, \xi)\right)  \right| \notag \\
& \leq \left| \X(\xi) - \X(\text{Re}\, \xi) \right| \notag \\
& \leq \frac{1}{4\gamma_j} \sup_{\xi \in \Xi_{\frac{1}{4\gamma_j}}} \left| \partial_{\xi} \X(\xi)  \right| \notag \\
& \leq \frac{1}{4\gamma_j} \sup_{\xi \in \Xi_{\frac{1}{4\gamma_j}}} \sum_q |q||\x(q)|e^{iq\cdot \xi} \notag \\
& \leq \frac{1}{4\gamma_j} \sum_q |q||\x(q)|e^{|q|\frac{1}{4\gamma_j}} \notag \\
& \leq \frac{1}{4\gamma_j} \label{stay}
\end{align}
using \eqref{xdecay} for $\varepsilon$ (i.e. $|\lambda|$) small enough; hence from the Cauchy estimates for analytic functions we get
\begin{align}
\Vert V^j_{n+1}(\theta + X(\theta)) \Vert_{\mathcal{L}(\C^{2d}, \ldots , \C^{2d}; \C^{2d})} \leq C n! \left(2\gamma_j\right)^{n} \quad \textrm{for some} \:  C \in \R
\end{align}
and finally using Cauchy Theorem we have for all $\eta \in \R$ such that $|\eta| \leq \frac{1}{4\gamma_j}$
\begin{align}
& \left|v^j_{n+1}(q;x) (Y_1, \ldots , Y_n)\right| \! = \notag \\
& \quad = \! \left| \frac{1}{(2\pi)^d} \int_{\T^d} V^j_{n+1}(\theta \!+\! i\eta \!+\! \X(\theta+ i\eta))(Y_1, \ldots , Y_n)e^{iq\cdot(\theta + i\eta)}d\theta \right| \notag \\
& \quad \leq \frac{1}{(2\pi)^d} \int_{\T^d}\left| V^j_{n+1}(\theta + i\eta + \X(\theta+ i\eta)) (Y_1, \ldots , Y_n) \right| e^{-q\cdot\eta} \notag \\
& \quad \leq C n! \left(2\gamma_j\right)^{n}  e^{-q\cdot\eta} |Y_1| \cdots |Y_n|
\end{align}
hence
\begin{align}
\Vert v_{n+1}^j(q;x)\Vert_{\mathcal{L}(\C^{d}, \ldots , \C^{d}; \C^{d})} \leq C n! \left(2\gamma_j \right)^{n}  e^{-q\cdot\eta}
\end{align}
and taking $\eta = \frac{1}{4\gamma_j} \frac{q}{|q|}$ we get \footnote{note that with such choice of $\eta$, because of \eqref{stay}, the complex number $\theta + i\eta + X(\theta + i\eta)$ belongs to the analyticity strip of the integrand function}
\begin{align}
\sum_{q \in \Z^d} e^{\sigma|q|} \Vert v_{n+1}^j(q;x)\Vert_{\mathcal{L}(\C^{d}, \ldots , \C^{d}; \C^{d})} \leq  \underbrace{C \sum_{q \in \Z^d} e^{(\sigma - \frac{1}{4\gamma_j})|q|}}_{:= b < \infty} n! (2\gamma_j)^{n} 
\end{align}
for all $0< \sigma < \frac{1}{5\gamma_j}$.

$\hfill \Box$

\subsection{Proof of Proposition \ref{start}}\label{Appstart}
Let us set
\begin{align}
w_0^{j(n)}(\x; q,q_1,\ldots,q_n) \equiv \frac{1}{n!}v_{n+1}^j(\x ; q-\sum_j q_j) 
\end{align}
inserting the Fourier expansion of $Y$, we can compute
\begin{align}
& \sum_{q \in \Z^d}|\w^j_{0\beta}(y ; q)| = \sum_{q \in \Z^d} |\tau_{\beta}w_0^j (\x + \tau_{-\beta}y;q) - \tau_{\beta}w_0^{j-1}(\x;q)| \notag \\
& = |\lambda| \sum_{n=0}^{\infty} \sum_{q,q_1,\ldots,q_n} \Big| e^{i \beta \cdot (q - \sum q_j)} w_0^{j(n)}(\x; q,q_1,\ldots,q_n)(y(q_1),\ldots,y(q_n)) \notag \\
& \quad - \sum_{q} e^{i \beta \cdot q}w_0^{j-1}(\x;q) \Big| \notag \\
& = |\lambda| \sum_{n=1}^{\infty} \sum_{q,q_1,\ldots,q_n} \Big| e^{i \beta \cdot (q - \sum q_j)} w_0^{j(n)}(\x; q,q_1,\ldots,q_n)(y(q_1),\ldots,y(q_n)) \notag \\
& \quad + \sum_{q} \w_{0\beta}^j(0;q) \Big| \notag \\
\end{align}
now \eqref{due} follows immediately from Lemma \ref{aprioridecay}, and \eqref{tre} follows from Lemma \ref{aprioridecay} and $\Vert y \Vert \leq \aalpha_j^{\frac{2}{3}\ell}$

To prove \eqref{uno},  for $|\eta| \leq \frac{1}{4\gamma_j}$, we use \eqref{stay}, the hypotheses on $V$ of Theorem \ref{KAM} and \eqref{constants} to get
\begin{align}
& \left| \partial V^j(\theta + i\eta+ \X(\theta + i\eta)) - \partial V^{j-1}(\theta + i\eta + \X(\theta + i\eta)) \right| \notag \\
& = \left| \sum_{\gamma_{j-1} < |q| \leq \gamma_j} q v(q)e^{iq\cdot(\theta + i\eta + \X(\theta + i\eta))} \right|\notag \\
& \leq \sum_{\gamma_{j-1} < |q| \leq \gamma_j}|q| |v(q)| e^{|q|\frac{1}{2\gamma_j}} \notag \\
& \leq \frac{1}{\gamma_{j-1}^{\ell}} \sum_{\gamma_{j-1} < |q| \leq \gamma_j}|q|^{\ell+1} v(q)e^{|q|\frac{1}{2 \gamma_j }} \leq C \frac{1}{\gamma_{j-1}^{\ell}}. \label{Del}
\end{align}

Then we choose $|\eta| =  \frac{1}{4\gamma_j}\frac{q}{|q|}$ and use \eqref{Del} to proceed as in Lemma \ref{aprioridecay} in order to get
\begin{align}
& | \w^j_{\beta 0} (0;q) | =  \left|e^{i \beta \cdot q} \left(w_0^j(\x;q) - w_0^{j-1}(\x;q)\right) \right| \notag \\ 
& \leq  e^{|\text{Im}\, \beta| |q|}\frac{\lambda}{(2\pi)^d} \int_{\T^d} \left| (V^j- V^{j-1})(\theta + i\eta + \X(\theta + i\eta)) \right| e^{iq\cdot(\theta + i\eta)} d\theta \notag \\
& \leq |\lambda| C \frac{1}{\gamma_{j-1}^{\ell}}  e^{(|\text{Im}\, \beta| - \frac{1}{4\gamma_j}) |q|} \leq |\lambda| C \frac{(8\gamma_j)^{\ell/3}}{\gamma_{j-1}^{\ell}|q|^{\ell/3}} \leq \varepsilon \frac{\aalpha_j^{\frac{2}{3}\ell}}{|q|^{\ell/3}}
\end{align}
for all $|\text{Im}\, \beta| < \frac{1}{8\gamma_j}  = \bar \alpha_j$.

Finally, in view of \eqref{sblit} we combine \eqref{uno}, \eqref{due}, \eqref{tre} and for $\ell$ large enough we obtain \eqref{zero}. This concludes the proof of the Proposition.

$\hfill \Box$

\subsection{Proof of Proposition \ref{easy}}\label{Appeasy}

Consider the fixed point equation \eqref{titcumulative} and write it as $w = \mathcal{F}(w)$, for $w = \w_{0\beta}^j$ and
\begin{align}
\mathcal{F}(w)(y) = \w_{0\beta}^j(y + \Gamma_{<n}w(y)).
\end{align}
Let
\begin{align}
 \mathcal{B}_n^j = \left\{w \in H^{\infty}(B(\aalpha_j^{\frac{2}{3}\ell}r^n) , \HH) \; | \; \Vert w \Vert_{\mathcal{B}_n^j} \equiv \sup_{y \in B(\aalpha_j^{\frac{2}{3}\ell}r^n)} \Vert w(y) \Vert \leq C_{d,\ell} \aalpha_j^{\frac{2}{3}\ell} |\lambda| \right\},
\end{align}
where $C_{d, \ell}$ is as in Prop. \ref{start}. Let us choose $\lambda_n$ such that $C\eta^{-2n}C_{d,\ell}\lambda_n \leq r^n$ for all $n$, with $C$ as in Lemma \ref{cutoff}. It follows from the latter that for $w \in \mathcal{B}_n^j$ and $y \in B(\aalpha_j^{\frac{2}{3}\ell} r^n) \subset \HH$,
\begin{align}
\Vert y + \Gamma_{<n}w(y) \Vert \leq \aalpha_j^{\frac{2}{3}\ell}r^n + C \eta^{-2n}C_{d,\ell} \alpha_j^{\frac{2}{3}\ell} |\lambda| \leq 2\aalpha_j^{\frac{2}{3}\ell}r^n \leq \frac{1}{2}\aalpha_j^{\frac{2}{3}\ell}  \label{virtue},
\end{align}
so $\mathcal{F}(w)$ is defined in $B(\aalpha_j^{\frac{2}{3}\ell}r^n)$ and, by Proposition \ref{start},
\begin{align}
\Vert \mathcal{F}(w)\Vert_{\mathcal{B}_n^j} \leq C_{d,\ell}\alpha_j^{\frac{2}{3}\ell} |\lambda|.
\end{align}
Hence $\mathcal{F} : \mathcal{B}_n^j \rightarrow \mathcal{B}_n^j$. For $w_1 , w_2 \in \B_n^j$ use \eqref{Cauchy3} to conclude that
\begin{align}
\Vert \mathcal{F}(w_1) - \mathcal{F}(w_2) \Vert_{\mathcal{B}_n^j} & = \!\!\!\!\! \sup_{\Vert y \Vert \leq \aalpha_j^{\frac{2}{3}\ell} r^n} \Vert \w_{0\beta}^j(y + \Gamma_{<n}w_1(y)) - \w_{0\beta}^j(y + \Gamma_{<n}w_2(y))\Vert \notag \\
& \leq \frac{2}{\aalpha_j^{\frac{2}{3}\ell}}C_{d,\ell}\aalpha_j^{\frac{2}{3}\ell}|\lambda| C\eta^{-2n}\Vert w_1 - w_2 \Vert_{\mathcal{B}_n^j} \notag \\
& \leq 2 r^n \Vert w_1 - w_2 \Vert_{\mathcal{B}_n^j} \notag \\ 
& \leq \frac{1}{2} \Vert w_1 - w_2 \Vert_{\mathcal{B}_n^j},
\end{align} 
i.e. $\mathcal{F}$ is a contraction. It follows that \eqref{titcumulative} has a unique solution $\w_{n\beta}^j$ in $\mathcal{B}_n^j$ satisfying the bound \eqref{go}, which, besides, is analytic in $\lambda$ and $\beta$.

Consider now for $n \geq 2$ the map $\mathcal{F}'$:
\begin{align}
\mathcal{F}'(w)(y) = \w_{0\beta}(y + \Gamma_{n-1}\w_{n\beta}(y) + \Gamma_{<n-1}w(y)) \label{Fprime};
\end{align}
again $\mathcal{F}'$ is a contraction in $\mathcal{B}^j_n$ since, for $\Vert y \Vert \leq \aalpha_j^{\frac{2}{3}\ell}r^n$, we have
\begin{align}
\Vert y + \Gamma_{n-1}\w_{n\beta}(y) + \Gamma_{<n-1}w(y) \Vert \leq 3\aalpha_j^{\frac{2}{3}\ell} r^n \leq \frac{1}{2} \aalpha_j^{\frac{2}{3}\ell}
\end{align}
for $r$ sufficiently small. But from Eqs. \eqref{titcumulative} one deduces that  $\w_{n\beta}^j$ and $\w_{(n-1)\beta}^j \circ \left( 1 + \Gamma_{n-1}\w_{n\beta}^j \right)$, both in $\mathcal{B}_n^j$, are its fixed points (just plug them into \eqref{Fprime}), hence by uniqueness they have to coincide, and \eqref{iterationt} follows. 

By virtue of the estimate \eqref{virtue} and definition \eqref{fcumulative}, 
\begin{align}
\sup_{\Vert y \Vert \leq \aalpha_j^{\frac{2}{3}\ell}r_j^n} \Vert f_{n\beta}^j(y) \Vert = \sup_{\Vert y \Vert \leq \aalpha_j^{\frac{2}{3}\ell}r_j^n}  \Vert y + \Gamma_{<n}\w_{n\beta}^j(y) \Vert \leq 2\aalpha_j^{\frac{2}{3}\ell} r^n. 
\end{align}
The recursion \eqref{iterationf} follows easily from Eq. \eqref{iterationt}:
\begin{align}
 f^j_{n\beta}(y) & =  y + \Gamma_{<n}\w_{n\beta}^j(y) \notag \\
& =  y + \Gamma_{n-1}\w_{n\beta}^j(y) + \Gamma_{<n-1}\w_{n\beta}^j(y) \notag \\
& = y + \Gamma_{n-1}\w_{n\beta}^j(y) + \Gamma_{<n-1} \w_{(n-1)\beta}^j(y + \Gamma_{n-1}\w_{n\beta}^j(y)) \notag \\
& = f_{(n-1)\beta}^j(y + \Gamma_{n-1}\w^j_{n\beta}(y)).
\end{align}

$\hfill \Box$

\subsection{Derivation of the modified Ward identity}\label{AppWard}

We shall prove \eqref{basic}, starting with $n=0$,
\begin{align}
& \int_{\T^d}\W_0^{\gamma}(Y; \theta)d\theta = \! \lambda \!\! \int_{\T^d}(\partial_{\gamma}V)(\theta + X(\theta) + Y(\theta))d\theta - \!\!\! \int_{\T^d}(\partial_{\gamma}\widehat{V})(\theta + X(\theta))d\theta \notag \\
&=  \lambda\int_{\T^d} \partial_{\gamma}\left( V(\theta + X(\theta) + Y(\theta)) \right)d\theta \notag \\
& \quad -  \lambda \int_{\T^d} (\partial_{\alpha}V)(\theta + X(\theta) + Y(\theta))\left(\partial_{\gamma}Y^{\alpha}(\theta)+ \partial_{\gamma}X^{\alpha}(\theta)\right)d\theta \notag \\
& \quad + \lambda\int_{\T^d} \partial_{\gamma}\left( \widehat{V}(\theta + X(\theta)) \right)d\theta -  \lambda \int_{\T^d} (\partial_{\alpha}\widehat{V})(\theta + X(\theta))\partial_{\gamma}X^{\alpha}(\theta)d\theta.
\end{align}
The first and the third term in the right hand side vanish, and by integrating the second and the fourth term by parts we get
\begin{align}
 \int_{\T^d}\W_0^{\gamma}(Y; \theta)d\theta & =  -  \lambda \int_{\T^d} \partial_{\gamma}(\partial_{\alpha}V)(\theta + X(\theta) + Y(\theta))\left(Y^{\alpha}(\theta)+ X^{\alpha}(\theta)\right)d\theta \notag \\
& \quad - \lambda \int_{\T^d} \partial_{\gamma}(\partial_{\alpha}\widehat{V})(\theta + X(\theta))X^{\alpha}(\theta)d\theta.
\end{align}
Writing $(\partial_{\alpha}V)(\theta + \X(\theta) + Y(\theta)) = W_0^{\alpha}(X+Y ; \theta)$, we get :
\begin{align}
\int_{\T^d}\W_0^{\gamma}(Y; \theta)d\theta = \int_{\T^d} Y^{\alpha}(\theta)\partial_{\gamma}W_0^{\alpha}(X+Y ; \theta)d\theta + \int_{\T^d} X^{\alpha}(\theta)\partial_{\gamma}\W_0^{\alpha}(Y ; \theta)d\theta. \label{basic0}
\end{align}
that is \eqref{basic} for $n=0$, since $X(\theta) = G_0U_0(\X ; \theta)$.
To prove the claim for $n \geq 1$, we use the relation \eqref{Titcumulative}:

\begin{align}
&\int_{\T^d} \W_n^{\gamma}(Y,\theta)d\theta = \int_{\T^d} \W_0^{\gamma}(Y + \Gamma_{<n}\W_n(Y) ; \theta) d\theta \notag \\
&\overset{(*)}{=} \int_{\T^d} \left(Y + \Gamma_{<n}\W_n(Y) \right)^{\alpha}\partial_{\gamma}W_0^{\alpha}(X + Y + \Gamma_{<n}\W_n(Y) ; \theta) d\theta \notag \\
& \quad + \int_{\T^d}  X^{\alpha}\partial_{\gamma}\W_n^{\alpha}( Y; \theta)d\theta \notag \\
& \overset{(**)}{=} \int_{\T^d}   Y^{\alpha} \partial_{\gamma}W_0^{\alpha}(X + Y + \Gamma_{<n}\W_n(Y) ; \theta) d\theta \notag \\
& \quad + \int_{\T^d} \Gamma_{<n}W_0^{\alpha}(X+Y + \Gamma_{<n}\W_n(Y) ; \theta)\partial_{\gamma}W_0^{\alpha}(X+Y + \Gamma_{<n}\W_n(Y) ; \theta) d\theta \notag \\
& \quad - \int_{\T^d} \Gamma_{<n}U_0^{\alpha}(X; \theta)\partial_{\gamma}W_0^{\alpha}(X + Y + \Gamma_{<n}\W_n(Y) ; \theta) d\theta \notag \\
& \quad + \int_{\T^d}X^{\alpha}\partial_{\gamma}W_0^{\alpha}(X + Y + \Gamma_{<n}\W_n(Y) ; \theta) d\theta \notag \\
& \quad - \int_{\T^d}G_0U_0^{\alpha}(X;\theta)\partial_{\gamma}U_0^{\alpha}(X;\theta) \notag \\
& \overset{(***)}{=} \int_{\T^d}   Y^{\alpha} \partial_{\gamma}W_0^{\alpha}(X + Y + \Gamma_{<n}\W_n(Y) ; \theta) d\theta \notag \\
& \quad + \int_{\T^d} G_{n}U_0^{\alpha}(X; \theta)\partial_{\gamma}W_0^{\alpha}(X + Y + \Gamma_{<n}\W_n(Y) ; \theta) d\theta \notag \\
\end{align}
where (*) comes from \eqref{basic0}, (**) from $X\!= G_0U_0(X)$, finally (***) is obtained from  $\X - \Gamma_{<n}U_0(\X) \!= G_nU_0\X$ plus
\begin{align}
& \int_{\T^d}\!\!\! \Gamma_{<n}(\theta-\theta')W_0^{\alpha}(X\!\!+\!Y \!+ \!\Gamma_{<n}\W_n(Y) ; \theta)\partial_{\gamma}W_0^{\alpha}(X\!\!+\!Y\! + \!\Gamma_{<n}\W_n(Y) ; \theta') d\theta d\theta'  \notag \\
& \quad - \int_{\T^d}G_0(\theta-\theta')U_0^{\alpha}(X;\theta)\partial_{\gamma}U_0^{\alpha}(X;\theta')d\theta d\theta' = 0\label{hence} 
\end{align}
which is obtained performing two integrations by parts and using the symmetry of $\Gamma_{<n}$ and $G_0$; the latter shows that the l.h.s. in \eqref{hence}  is equal to its opposite, hence it vanishes.

\newpage
\begin{section}*{Acknowledgements}
First and foremost I would like to thank my supervisor Antti Kupiainen for introducing me to KAM theory and for teaching me most of what I know about it.
I would like to express sincere gratitude to my friend and collegue Mikko Stenlund for his companionship during my PhD years and for the fruitful discussions that have helped me improving the present work and my overall understanding of the subject.  
Also, I want to acknowledge the whole Mathematical Physics Group and the staff of the Department of Mathematics at the University of Helsinki.

Finally, I would like to thank the Academy of Finland for financial support.
\end{section}

\nocite{HZ, Arnold2, Dittrich, Kolmogorov1, Moser1, Moser2, Gallalindt, Gallatwist, Gallatwist2, Gentile1, Arnold1, Arnold2, Arnold3, Poincare, Pöschel1, Pöschel2, Pöschel3, Antti1, Chierchia1, Katok, Eliasson, Thirring, Salomon1}
\bibliographystyle{amsplain}

\begin{thebibliography}{10}

\bibitem{Arnold2}
V.I. Arnol'd, \emph{{P}roof of a theorem by {A}. {N}. {K}olmogorov on the
  invariance of quasi-periodic motions under small perturbations of the
  {H}amiltonian}, {R}uss. {M}ath. {S}urv. \textbf{18} (1963), no.~5, 9--36.

\bibitem{Arnold4}
\bysame, \emph{{S}mall denominators and problems of stability of motion in
  classical and celestial mechanics}, {R}uss. {M}ath. {S}urv. \textbf{18}
  (1963), no.~6, 9--36, [Corrigenda (in Russian): \textit{Uspekhi Math. Nauk.}
  \textbf{23} (1968) 216.].

\bibitem{Arnold1}
\bysame, \emph{{M}athematical {M}ethods of {C}lassical {M}echanics}, {I} ed.,
  Springer, 1978.

\bibitem{Arnold3}
\bysame, \emph{Geometrical {M}ethods in the {T}heory of {O}rdinary
  {D}ifferential {E}quations}, Springer-Verlag, New York, 1987.

\bibitem{Antti1}
J.~Bricmont, K.~Gaw\c edzki, and A.~Kupiainen, \emph{{KAM} {T}heorem and
  {Q}uantum {F}ield {T}heory}, Mathematical Physics Preprint Archive (1999).

\bibitem{Chierchia2}
  L.~Chierchia,
  \emph{K{A}{M} {L}ectures}, 
  Dynamical Systems. Part I: Hamiltonian Systems and Celestial Mechanics,
Pubblicazioni della Classe di Scienze, Scuola Normale Superiore, Pisa. Pagg. 1-56 (2003). 

\bibitem{Chierchia3}
  L.~Chierchia and C.Falcolini,
  \emph{Compensations in {S}mall {D}ivisor {P}roblems}, 
  Commun. math. phys. 1996, vol. 175, no1, pp. 135-160

\bibitem{Chierchia1}
L.~Chierchia and C.Falcolini, \emph{A {D}irect {P}roof of a {T}heorem by
  {K}olmogorov in {H}amiltonian {S}ystems}, Ann. Sc. Norm. Sup. Pisa, Serie IV, 1994.
  
\bibitem{Dittrich}
W.~Ditrich and M.Reuter, \emph{Classical and {Q}uantum {D}ynamics}.

\bibitem{Eliasson}
L.~H. Eliasson, \emph{Absolutely convergent series expansions for quasi
  periodic motions}, Reports Department of Math., Univ. of Stockholm, Sweden
  (1988), 1--31.

\bibitem{Feldman1}
J.~Feldman and E.~Trubowitz, \emph{Renormalization in classical mechanics and
  many body quantum field theory}, 1992.

\bibitem{Gallatwist2}
G.Gallavotti, \emph{{T}wistless {K}{A}{M} tori.}, {C}omm. {M}ath. {P}hys.
  \textbf{164} (1994), no.~1, 145--156.

\bibitem{Gallatwist}
\bysame, \emph{{T}wistless {K}{A}{M} tori, quasi flat homoclinic intersections,
  and other cancellations in the perturbation series of certain completely
  integrable {H}amiltonian systems. {A} review.}, {R}ev. {M}ath. {P}hys.
  \textbf{6} (1994), no.~3, 343--411.

\bibitem{Gallalindt}
\bysame, \emph{{L}indstedt series and {K}olmogorov theorem.}, Hamiltonian
  systems with three or more degrees of freedom (C.~Simó, ed.), NATO Science
  Series C: Math.Phys.Sci., vol. 533, Springer, first ed., June 1999,
  pp.~62--71.

\bibitem{GallaRG}
\bysame, \emph{Renormalization {G}roup in {S}tatistical {M}echanics and {M}echanics: gauge symmetries and vanishing beta functions}, 
{P}hysics {R}eports, \textbf{352}, 251-272 (2001)


\bibitem{Gentile1}
G.Gentile and V.Mastropietro, \emph{{M}ethods for the analysis of the lindstedt
  series for {K}{A}{M} tori and renormalizability in classical mechanics. {A}
  review with some applications}, {R}ev. {M}ath. {P}hys. \textbf{8} (1996),
  393--444.

\bibitem{HZ}
H.~Hofer and Eduard Zehnder, \emph{Symplectic {I}nvariants and {H}amiltonian
  {D}ynamics}, Birkhäuser Verlag, 1994.

\bibitem{Katok}
A.~Katok and B.~Hasselblatt, \emph{Introduction to the modern theory of
  dynamical systems}, Cambridge University Press, 1995.

\bibitem{Kolmogorov1}
A.~N. Kolmogorov, \emph{{O}n the conservation of conditionally periodic motions
  for a small change in {H}amilton's function}, {D}okl. {A}kad. {N}auk. {SSSR}
  \textbf{98} (1954), 525--530, in Russian; English translation in \textit{LNP}
  \textbf{93} (1979) 51-56.

\bibitem{Moser1}
J.~Moser, \emph{{O}n invariant curves of area-preserving mappings of an
  annulus}, Nachr. Akad. Wiss. Göttingen, Math-phys. \textbf{Kl. II} (1962),
  no.~1, 1--20.

\bibitem{Moser2}
\bysame, \emph{{C}onvergent series expansions for quasi-periodic motions},
  Math. Ann. \textbf{169} (1967), 136--176.

\bibitem{Poincare}
H.~Poincar\'e, \emph{Les m\'ethodes nouvelles de la m\'ecanique c\'eleste},
  vol. 1-3, Paris: Gauthier-Villars, 1892/1893/1899.

\bibitem{Pöschel1}
J.~Pöschel, \emph{{Ü}ber invariante tori in differenzierbaren {H}amiltonschen
  {S}ystemen}, {B}onn. {M}ath. {S}chr. \textbf{120} (1980), 1--103.

\bibitem{Pöschel2}
\bysame, \emph{{I}ntegrability of {H}amiltonian systems on {C}antor sets},
  {P}ure {A}ppl. {M}ath. \textbf{35} (1982), 653--695.

\bibitem{Pöschel3}
\bysame, \emph{{A} {L}ecture on the {C}lassical {K}{A}{M} {T}heorem}, {P}roc.
  {S}ymp. {P}ure {M}ath. \textbf{69} (2001), 707--732.

\bibitem{Salomon1}
Dietmar~A. Salamon, \emph{{T}he {K}olmogorov-{A}rnold-{M}oser theorem},
  {M}athematical {P}hysics {E}lectronic {J}ournal (2004).

\bibitem{Thirring}
W~Thirring, \emph{Classical {D}ynamical {S}ystems}, Springer-Verlag, 1978.

\end{thebibliography}

\providecommand{\bysame}{\leavevmode\hbox to3em{\hrulefill}\thinspace}
\providecommand{\MR}{\relax\ifhmode\unskip\space\fi MR }
\providecommand{\MRhref}[2]{%
  \href{http://www.ams.org/mathscinet-getitem?mr=#1}{#2}
}
\providecommand{\href}[2]{#2}

\end{document}